\documentstyle[aaspp4,flushrt]{article}
\input psfig.sty

\makeatletter

\@ifundefined{chapter}{\def\thebibliography#1{\section*{References\@mkboth
  {REFERENCES}{REFERENCES}}\list
  {\relax}{\setlength{\labelsep}{0em}
        \setlength{\itemindent}{-\bibhang}
        \setlength{\itemsep}{0pt}
        \setlength{\parsep}{0pt}
        \setlength{\leftmargin}{\bibhang}}
    \def\newblock{\hskip .11em plus .33em minus .07em}
    \sloppy\clubpenalty4000\widowpenalty4000
    \sfcode`\.=1000\relax}}%
{\def\thebibliography#1{\chapter*{Bibliography\@mkboth
  {BIBLIOGRAPHY}{BIBLIOGRAPHY}}\list
  {\relax}{\setlength{\labelsep}{0em}
        \setlength{\itemindent}{-\bibhang}
        \setlength{\itemsep}{0pt}
        \setlength{\parsep}{0pt}
        \setlength{\leftmargin}{\bibhang}}
    \def\newblock{\hskip .11em plus .33em minus .07em}
    \sloppy\clubpenalty4000\widowpenalty4000
    \sfcode`\.=1000\relax}}

\newlength{\bibhang}
\let\@internalcite\cite
\def\cite{\let\@citeleft(\let\@citeright)%
    \@ifstar{\citeyear}{\citefull}}
\def\citenp{\let\@citeleft\relax\let\@citeright\relax
    \@ifstar{\citeyear}{\citefull}}
\def\citefull{\def\astroncite##1##2{##1~##2}\@internalcite}
\def\citeyear{\def\astroncite##1##2{##2}\@internalcite}

\def\@citex[#1]#2{\if@filesw\immediate\write\@auxout{\string\citation{#2}}\fi
  \def\@citea{}\@cite{\@for\@citeb:=#2\do
    {\@citea\def\@citea{; }\@ifundefined
       {b@\@citeb}{{\bf ?}\@warning
       {Citation `\@citeb' on page \thepage \space undefined}}%
{\csname b@\@citeb\endcsname}}}{#1}}

\def\@cite#1#2{\@citeleft#1\if@tempswa , #2\fi\@citeright}
\def\@biblabel#1{}

\makeatother

\def\gsim{\;\rlap{\lower 2.5pt
 \hbox{$\sim$}}\raise 1.5pt\hbox{$>$}\;}
\def\lsim{\;\rlap{\lower 2.5pt
   \hbox{$\sim$}}\raise 1.5pt\hbox{$<$}\;}

\def\angle0{\mbox{\boldmath$\theta_{\rm planet}$}}

\def\Deltabeta{\mbox{\boldmath$\Deltabeta$}}

\begin{document}

\title{Towards Constraints on Dark Energy from Absorption Spectra of Close Quasar Pairs}
\author{Adam Lidz$^{a}$, Lam Hui$^{a,b,c}$, Arlin P.S. Crotts$^{d}$, and
Matias Zaldarriaga$^{e}$}

\affil{$^{a}$ Department of Physics, Columbia University, 538 West 120th Street,
New York, NY 10027\\   
$^{b}$ Theoretical Astrophysics, Fermi National Accelerator Laboratory,
P.O. Box 500, Batavia, IL 60510\\
$^{c}$ Department of Astronomy and Astrophysics, University of Chicago,
IL 60637\\
$^{d}$ Department of Astronomy, Columbia University, 538 West 120th Street,
New York, NY 10027\\
$^{e}$ Departments of Astronomy and Physics, Harvard University, 60 Garden 
Street, Cambridge, MA 02138\\
{\tt lidz@astro.columbia.edu, lhui@fnal.gov, 
arlin@astro.columbia.edu, mzaldarriaga@cfa.harvard.edu
}}

\begin{abstract}
A comparison between the line of sight power spectrum (the auto spectrum) 
of absorption in the Lyman-alpha forest and the cross power
spectrum (the cross spectrum) between the absorption in 
neighboring lines of sight offers an evolution-free means to 
constrain the cosmological constant, or dark energy. Using cosmological 
simulations, we consider a maximum likelihood method to obtain constraints 
from this comparison. In our method, measurements of the auto and cross 
spectra from observations are compared with those from a multi-parameter 
grid of simulated models of the intergalactic medium (IGM). We then 
marginalize over nuisance parameters to obtain constraints on the 
cosmological constant.
Redshift space distortions due to peculiar velocities and thermal broadening,
a potential difficulty in applying this test, are explicitly modeled in our 
simulations. To illustrate our method, we measure the cross spectrum
from a new sample of five close quasar pairs, with separations of 0.5 to 3 
arcmin. Using published measurements of the auto spectrum and our 
measurements of the cross spectrum, we attempt to obtain a 
constraint on $\Omega_\Lambda$, but find only weak constraints. An 
Einstein-de-Sitter cosmology is, however, disfavored by the data at 
a $\sim 2\sigma$ confidence level. We consider the power of future 
observations, paying particular attention to the effects of spectral 
resolution and shot-noise. We find that $\sim 50$ moderate resolution, 
FWHM $\sim 150$ km/s, fully-overlapping, close separation pairs 
(with $\Delta \theta \sim 30''-120''$) should allow a ($2\sigma$) 
constraint on $\Omega_\Lambda$ at the level of $15 \%$, if other 
modeling parameters are well determined through other means. We find 
that there is a sizeable gain from observing very close, 
$\Delta \theta \sim 30''$, separation pairs provided they are observed with 
high spectral resolution. A sample of $\sim 10$ such pairs would yield 
similar constraints to the $50$ moderate resolution pairs mentioned above.
\end{abstract}


\section{Introduction}
\label{intro}

In the last several years, much work has been done on the measurement of
the flux/transmission auto power spectrum from the Lyman-alpha forest, and the
derivation of cosmological constraints from it 
(e.g. Croft et al. 1998,1999,2002, Hui 1999, McDonald et al. 2000, 
Zaldarriaga, Hui \& Tegmark 2001, 
Zaldarriaga, Scoccimarro \& Hui, 2003, Gnedin \& Hamilton 2002). 
Motivated by these developments, it is natural to extend observations 
and simulations to compare the cross-correlation, or 
in Fourier space the cross spectrum, between the absorption in 
neighboring lines of sight. 
Measuring the flux cross spectrum is a more
ambitious project observationally than measuring the auto spectrum 
because extracting a good signal requires Ly$\alpha$ 
forest spectra of many close quasar pairs, which are rare.
However, good observations of the Ly$\alpha$ forest
in a handful of close quasar pairs have already existed for several years 
(Bechtold et al. 1994, Crotts \& Fang 1998, Petitjean et al. 1998), 
and yielded information
on the size and shape of Ly$\alpha$ absorbers.
Theoretically, simulating the cross-correlation is challenging since it 
depends more sensitively on redshift distortions than the auto spectrum, 
the detailed modeling of which
requires large volume, high resolution simulations. Although difficult to
obtain, the information contained in the cross-correlation signal is rich.
In particular, using the auto spectrum of the
absorption in the Ly$\alpha$ forest and the cross spectrum one can
constrain the cosmological constant or dark energy through
a version of the Alcock-Paczy\'nski (AP) test (Alcock \& Paczy\'nski 1979,
McDonald \& Miralda-Escud\'e 1999, Hui, Stebbins \& Burles 1999,
McDonald 2003, Lin \& Norman 2002, Rollinde et al. 2003). 
In spite of this, little work has been done 
in comparing pair spectra with simulations using
the modern picture of the Ly$\alpha$ forest and focusing on continuous-field
statistics. (See however Rollinde et al. 2003.) 

As a step in this direction, this paper 
aims to measure the cross spectrum from data and from 
simulations. Our present data set, consisting of five pairs
with $z \gtrsim 2.2$, includes two new pairs and a triplet from Crotts
\& Fang (1998). 
From our measurements of the cross spectrum,
which have large error bars owing to the small number of pairs in our
current sample, and using the precise measurements of the auto spectrum
from Croft et al. (2002), we attempt to obtain a constraint on 
the cosmological constant. 

It is appropriate to ask: why carry out this exercise if we already
know the value of the cosmological constant to fair accuracy
from supernova measurements (Riess et al. 1998, Perlmutter et al. 1999), especially
when used in conjunction with the microwave background anisotropies (Spergel et al. 2003)? The answer lies partly in the elegance of the 
Alcock-Paczy\'nski test, in that
it does not require a standard candle, and is therefore free of assumptions
about evolution. The application to the Ly$\alpha$ forest of course
suffers from systematics of its own. It is the cross-checks provided by
different methods that give us confidence in our current picture of the universe. The present paper should be viewed as a warm-up exercise -- 
as quasar pair samples increase in size, the method described here might
prove to give competitive constraints on the cosmological constant/dark energy,
and as will be discussed below, the resulting constraints will be
nicely complementary to those from other methods.

The paper is organized as follows. 
In \S \ref{AP} we briefly review
the AP test, its application using the Ly$\alpha$ forest and its
sensitivity to cosmological parameters, especially dark energy. In \S \ref{sim_stuff} we
describe the statistics that we use in our implementation of the
AP test and describe how we measure them from simulations.
In \S \ref{redshift distort} we describe and quantify the effect
of redshift distortions on the three-dimensional flux power spectrum
by measurements from simulations. In \S \ref{data} we present measurements
of the cross spectrum from a small sample of quasar pairs and attempt
to place a constraint on the cosmological constant from the measurements.
The constraint comes from comparing simulated auto and cross spectra 
from a grid of models with the observed auto and cross spectra. 
The grid of models examined here
is rather small, but our constraints are correspondingly weak.
In \S \ref{predictions} we consider the possibilities for constraining
dark energy with future observations and consider the dependence
on spectral resolution and shot noise. This section should be helpful
in planning future observations. In \S \ref{conclusions} we conclude.
In Appendix A we include some details concerning the simulations
we run and show resolution and boxsize tests.

\section{The AP Test}
\label{AP}

In this section we briefly review the AP test and its
application in the Ly$\alpha$ forest.
Alcock \& Paczy\'nski (1979) considered an astrophysical 
object, with redshift extent $\Delta z$ and angular extent $\Delta \theta$,
and pointed out that the object's inferred shape depends on cosmology.
To be specific, the dependence on cosmology is as follows.
The length of the hypothetical object along the line of sight 
in velocity units 
is $u_\parallel = {{c \Delta z} \over {1+z}}$, and its perpendicular 
extent in velocity units is 
$u_\perp = {H(z) \over 1+z} D_A (z) \Delta \theta$.
Here $z$ is the mean redshift of the object, and $H(z)$ and $D_A(z)$ are 
respectively the Hubble parameter and the co-moving angular diameter 
distance at the mean redshift. The radiation-free Hubble parameter is 
$H(z) = H_0 \left[\Omega_m (1+z)^3 + \Omega_k (1+z)^2 + 
\Omega_\Lambda \right]^{1 \over 2}$, where $H_0$ is the present 
day Hubble parameter, and $\Omega_m, \Omega_k, 
\Omega_\Lambda$ are the present day matter, curvature and cosmological 
constant contributions to the energy density. The curvature density is 
related to the matter and cosmological constant densities by 
$\Omega_k = \vert 1 - \Omega_m - \Omega_\Lambda \vert$. The co-moving 
angular diameter distance, (as opposed to the proper angular diameter
distance), is $D_A (z) = {c \over H_0} 
{1 \over \sqrt{\Omega_k}} S_k \left[\sqrt{\Omega_k} \int_0^z 
{dz^\prime \over H(z^\prime)/H_0}\right]$. The function $S_k (x)$ that appears
in the angular diameter distance is 
$S_k (x) = \sin(x)$ for a closed universe, 
$S_k (x) = \sinh(x)$ for an open 
universe and $S_k (x) = x$ for a flat universe, in which case $D_A (z)$ is 
just the co-moving distance out to redshift z. 

\begin{figure}[htb]
\centerline{\psfig{figure=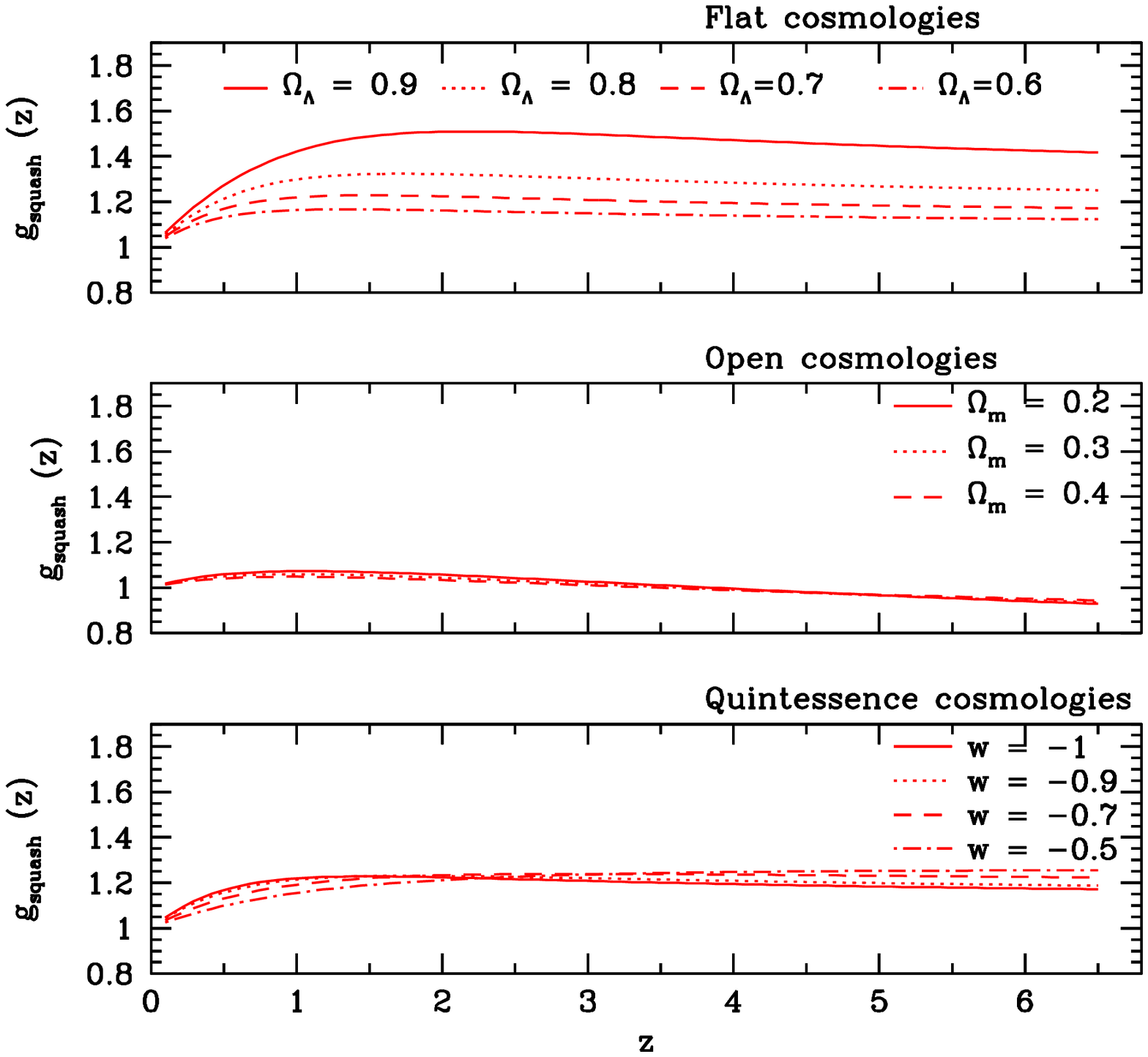,height=7.0in}}
\caption{\label{squash}
The top panel shows the squashing factor, $g_{\rm squash} (z)$, 
defined in \S \ref{AP}, as a function of redshift for various 
flat, Lambda cosmologies. The middle panel shows the same factor for 
open universes. The bottom panel shows the squashing factor for 
flat quintessence models
with $\Omega_m=0.3$. The equation of state of the quintessence field 
is assumed to be constant as a function of redshift. 
In each panel, the squashing factor is defined relative to the EDS
case; it is the squashing an observer converting angles and redshifts
to distances assuming an EDS cosmology would measure in a given ``true'' 
cosmology. 
}
\end{figure}

The ratio,
$u_\parallel \over u_\perp$, is a measure of the object's shape, equal to
one for spherical objects. The cosmological dependence of this ratio means 
that observers converting angles and redshifts to
distances assuming different cosmologies will reach different conclusions 
about the object's shape. One can assume any possible value of the
cosmological parameters $\Omega_m$, $\Omega_\Lambda$; if the assumed
parameters differ from the true cosmology, an object's inferred shape will 
be distorted from its true shape.
In particular we consider the distortion an observer assuming an
Einstein-de-Sitter (EDS)
cosmology measures in a universe with a cosmological constant (a Lambda
cosmology), or any other postulated ``true''
cosmology. We define a squashing factor, $g_{\rm squash} (z)$, to
quantify the distortion in the object's shape (see e.g. Ballinger et al.
1997),
\begin{equation}
g_{\rm squash} (z) = 
\left({{u^\prime_\perp} \over {u^\prime_\parallel}}\right) 
\left({{u_\parallel} \over {u_\perp}}\right)
= {{H^\prime(z) D^\prime_A(z)} \over {H(z) D_A (z)}}
\label{fsquash}
\end{equation} 
The primed variables represent coordinates in the EDS cosmology 
and the unprimed variables denote coordinates
in the Lambda (or other postulated ``true'') cosmology. 
This ratio is larger than one in the case of a Lambda cosmology;
${{H^\prime(z) D^\prime_A(z)} \over {H(z) D_A (z)}} > 1$. The object appears 
squashed along the line of sight to the observer converting redshifts and
angles to distances assuming an EDS cosmology; i.e., a spherical object with 
true dimensions $u_\parallel =  u_\perp$, seems squashed by a 
factor $g_{\rm squash}$ so that $u^\prime_\perp > u^\prime_\parallel$.
In figure (\ref{squash}) we plot the squashing factor, $g_{\rm squash}(z)$,
as a function of redshift for several different cosmological models.
In each case, the squashing factor shown is the squashing an EDS observer
measures in a given ``true'' cosmology.    
In the top panel we show the squashing factor for
several different values of $\Omega_{\Lambda}$, assuming a flat universe. 
In the middle panel we show the squashing factor assuming an open universe with
no cosmological constant. The squashing factors in the open universe, zero
cosmological constant models are close to unity, the squashing effect is 
always $\lesssim 10 \%$ and typically only a $\sim 5 \%$ effect in 
these models. 
As we can see in the top panel of the figure, the squashing effect
is rather sensitive, however, to the presence of a cosmological constant.  
The squashing factor in the flat, Lambda universes shown here
rises quickly with redshift, but plateaus around $z \sim 1$. 
After reaching this plateau, the squashing effect is only a 
$\sim 20 \%$ effect for an 
$\Omega_\Lambda = 0.7$ universe, but is a $\sim 50 \%$ effect for an 
$\Omega_\Lambda=0.9$ flat universe. The plateau effect arises because the
squashing is sensitive to the presence of a cosmological constant, which is 
small at high redshift. At redshifts beyond the plateau, 
the AP test should be particularly good at constraining the existence of 
a very large cosmological constant.

We also consider the sensitivity of the AP test to the equation of state
of a dark energy, or quintessence, component to the energy density
(Hui, Stebbins \& Burles 1999).
The equation of state of the quintessence component is parameterized 
by $p=w \rho_Q$, 
where $p$ is the (negative) pressure, and $\rho_Q$ is the 
component's energy density. 
Here we consider the equation of state to be constant as a function
of redshift, in which case $\rho_Q = \Omega_Q (1+z)^{3 (1+w)}$.
In the bottom panel of figure (\ref{squash}) we plot the squashing factor
as a function of redshift for a few different quintessence cosmologies.
At redshifts near $z=3$, where one can apply the AP test to the 
Ly$\alpha$ forest, the squashing factor is extremely similar for
all of the quintessence cosmologies considered. This is in accord
with McDonald (2003), (see also Kujat et al. 2002), 
who notes that this insensitivity to $w$ 
makes the Ly$\alpha$ forest a particularly good probe of $\Omega_m$
since constraints obtained on $\Omega_m$ are then insensitive to
prior assumptions regarding $w$. The squashing factor at  
this redshift is more sensitive to the equation of state of dark energy, $w$, 
if $w(z \sim 3) \gtrsim -0.4$, however (McDonald 2003).
For instance, as the equation of state of the dark energy approaches the
limiting case of $w=-1/3$ the effect approaches that of an open cosmology,
which has a significantly different squashing factor than a universe with
a cosmological constant ($w=-1$).\footnote{As we discuss below, the 
expected signal from the
Ly$\alpha$ AP test also depends on the effect of redshift distortions. 
The equation of state $w$ does affect redshift distortions,
but only to a small degree, as long as $w \lesssim -0.4$.}
At lower and higher redshifts, the squashing factor is more sensitive
to the equation of state, $w$. At low redshift the difference between
squashing factors for the different quintessence cosmologies peaks
near $z \sim 0.5$. The difference between squashing factors in a $w=-0.5$
cosmology and in a $w=-1$ cosmology is still only $\sim 6\%$ near this
redshift, but the clustering of low/intermediate 
redshift objects is certainly a better
probe of $w$ than the Ly$\alpha$ forest at $z \sim 3$. 
For instance, Matsubara \& Szalay (2002)
suggest applying the AP test using the clustering of luminous red 
galaxies, as a good low/intermediate 
redshift sample, to constrain $w$. At any rate, we now have an idea of the
size of the geometric distortion in different cosmologies, and proceed
to discuss some complications in applying the AP test with real 'astrophysical
objects'.

To apply this test in the pure form mentioned above, one needs a 
spherical
object or an object of other known shape, expanding with the Hubble flow. 
The three-dimensional power spectrum of galaxy or quasar clustering, 
$P(k)$, for instance, 
could provide such an object since it should be spherical by 
the cosmological principle if measured in real space (see e.g., Phillips 1994, 
Matsubara \& Suto 1996, Ballinger,Peacock \& Heavens 1997, 
Popowski et al. 1998, Outram et al. 2001, Hoyle et al. 2002, Seo \&
Eisenstein 2003. See also
Ryden 1995 and Ryden \& Melott 1996 for an application using the shape
of voids). This quantity is not directly accessible to observations, 
however, because measurements are
performed in redshift space rather than in real space; one measures redshifts
rather than absolute distances along the line of sight.\footnote{Of course
one can measure the real space power spectrum by considering only 
transverse modes, i.e. $P_r(k)=P_s(k_\parallel=0,k_\perp=k)$ (see e.g.
Hamilton \& Tegmark 2002, Zehavi et al. 2003), but this
is not helpful for the AP test where one wants to compare clustering
along the line of sight to that across the lines of sight.}
The shape of the
observed three-dimensional power spectrum is thereby distorted by the 
presence of peculiar velocities which affect distances inferred 
along the line of sight, but
not across the line of sight. On large scales, as overdense regions
undergo gravitational infall,
the clustering pattern is squashed along the line of sight 
by these
redshift-space distortions, mimicking the effect of a cosmological constant.
On small scales, redshift distortions elongate structure along the
line of sight in a ``finger-of-god'' effect.
To use the AP test
one must accurately model the effect of the redshift distortions. In the case
of galaxy or quasar clustering, this requires knowledge of the biasing 
between the dark matter clustering and the galaxy/quasar clustering.
The underlying physics of galaxy biasing is somewhat uncertain, 
making a first principles prediction
of the bias difficult, forcing one to determine the biasing in other ways.
The use of the galaxy bispectrum is one such promising method (Fry 1994, 
Scoccimarro et al. 2001, Verde et al. 2002), but
it is as yet unclear whether the bias obtained in this
way can be used in conjunction with the usual linear Kaiser formula 
(Kaiser 1987)
to describe redshift distortions to sufficient accuracy at
mildly nonlinear scales (Scoccimarro \& Sheth 2002).
This makes application of the AP test to galaxy surveys non-trivial,
because one must disentangle the degenerate effects of redshift distortions,
which depend on the biasing, and the cosmological distortion. 

The proposed versions of the AP test using the Ly$\alpha$ forest 
at least 
partly circumvent this difficulty (McDonald \& Miralda-Escud\'e 1999, 
Hui, Stebbins \& Burles 1999). In these versions of the AP test, one uses 
the power spectrum of fluctuations in absorption in the Ly$\alpha$ 
forest as the 'astrophysical object' for applying the test. In the
case of the forest, there is also a biasing that affects the redshift 
distortions, but the relevant biasing is between the fluctuations
in the Ly$\alpha$ absorption and the dark matter fluctuations.
The advantage provided by the forest is that the physics of the 
absorbing gas is relatively simple, unlike the physics involved in galaxy biasing. 
The relevant physics is primarily just that of gas in 
photoionization equilibrium with an external radiation field. 
On large scales the hydrogen gas distribution 
follows the dark matter distribution -- i.e. gravity-driven, and on small scales it is 
Jeans pressure-smoothed (see e.g. Cen et al. 1994, Zhang et al. 1995, Hernquist et al. 1996, 
Miralda-Escud\'e 1996, Muecket et al. 1996, 
Bi \& Davidsen 1997, Bond \& Wadsley 1997, Hui et al. 1997, Croft
et al. 1998, Theuns et al. 1999, Bryan et al. 1999, Nusser \& Haehnelt 1999). 
The hope is that one can thereby 
use numerical simulations to model this simple physics, and
understand the biasing between the flux power spectrum and the mass power 
spectrum. In this way we might model the effect of redshift distortions 
from first principles and thereby isolate the cosmological distortion.
One might worry that the relevant physics is actually not so simple and that
the small scale structure of the absorbing gas is 
modified by hydrodynamic effects, such as massive outflows from galaxies (Adelberger et al. 2003), that are not
modeled properly, or at all, in simulations.
There is direct observational evidence that indicates that
this is not the case, at least near $z \sim 3$ 
for the low column density gas responsible for most
of the absorption in the Ly$\alpha$ forest. Rauch et al. (2001), 
by directly measuring
the cross-correlation between lines of sight towards a
lensed quasar, find, in fact, that the density field is 
quite smooth on small scales, arguing against 
a turbulent IGM constantly stirred up by massive outflows.
One might also worry about the effect of a fluctuating ionizing background, but
at $z \sim 3$, this has been shown to be negligible on relevant scales
of interest (Croft et al. 1999, Meiksin \& White 2003). A further possible 
systematic effect is from spatial fluctuations in the temperature of the 
IGM (Zaldarriaga 2002), expected if HeII is reionized 
near $z \sim 3$ (Theuns et al. 2002). 
Another line of evidence, however, that supports the simple 
gravitational instability
picture of the forest comes from higher order statistics -- gravity
predicts a unique hierarchy of correlations, which so far seems to
be consistent with data (either from bispectrum type measurements (Zaldarriaga et al. 2001b;
see however Viel et al. 2003), or
the one point probability distribution of flux (PDF) (e.g. Gaztanaga \& Croft 1999, McDonald et al. 2000a)).
While increasing amount of data will test these assumptions more precisely, it
is worthwhile to push theoretical and observational efforts forward 
to obtain further cosmological constraints using, for instance, the AP test.

One other important 
difference with the case of galaxy clustering is that in the Ly$\alpha$ forest
one does not directly measure the full three-dimensional flux power spectrum.
Instead one measures the power spectrum of flux fluctuations along a line
of sight (the auto spectrum) and the cross-correlation of power between 
two neighboring lines of sight (the cross spectrum). 
An observer measuring a given auto spectrum will infer the wrong
cross spectrum if the observer assumes an incorrect cosmology in converting
from the observed angular separation between the lines of sight to
their transverse velocity separation. If the true cosmology is a Lambda cosmology, an observer converting between angles and distances assuming an EDS 
cosmology will infer too large a transverse velocity separation between the 
two lines of sight and predict too weak a cross correlation.

\section{Simulation measurements of the flux power spectra}
\label{sim_stuff}

Our basic strategy for using the AP test in the Ly$\alpha$ forest is 
to simultaneously generate auto spectra and cross spectra 
from simulations for a range of different models describing the IGM.
A model that fits the observed auto spectrum will only match the 
observed cross spectrum, 
measured at a given transverse separation, $\Delta \theta$, if the 
correct cosmology is assumed in converting from the observed $\Delta \theta$
to transverse velocity, $u_\perp (\Delta \theta)$. This, at least in 
principle,
should allow us to constrain the underlying cosmology of the universe. In this
section we define the relevant statistics and describe
how we measure them from simulations. (Hui 1999, Viel et al. 2002)

It is instructive
to describe the relationship of the auto spectrum and the cross spectrum
to the fully three-dimensional flux power spectrum. In the presence of redshift
distortions, the three-dimensional
power spectrum of the flux field\footnote{The flux power spectra 
considered in this paper are power spectra of
the random field $\delta_f = (f - \langle f \rangle)/\langle f \rangle$, where
$f$ is the transmitted flux and $\langle f \rangle$ is its mean.
Observationally this quantity is not sensitive to estimates of the absolute
continuum level, unlike the quantity $\tilde{\delta}_f = f - \langle f \rangle$, which is sometimes used. (Hui et al. 1999)}, $\delta_f$, depends on 
both the magnitude
of the three dimensional wave vector, $k$, and its line of sight component, 
$k_\parallel$, as we discuss in \S \ref{redshift distort}.
The auto spectrum, $P_{\rm 1d} (k_\parallel)$, is related to the full (not
directly observable) three-dimensional flux 
power spectrum, $P_F (k,\frac{k_\parallel}{k})$,  
by
\begin{equation}
P_{\rm 1d} (k_\parallel) = \int_{k_\parallel}^\infty \frac{d k^\prime}{2 \pi} 
k^\prime
P_F (k^\prime, \frac{k_\parallel}{k^\prime})
\label{pauto}    
\end{equation}
which is the general relationship between the three-dimensional power spectrum
of an azimuthally symmetric random field and the 
projected power along a line of sight (Kaiser \&
Peacock 1991). Generally instead of the auto spectrum we will plot the
dimensionless quantity, $k_\parallel P_{\rm 1d} (k_\parallel) /\pi$, 
which is the contribution per interval in ln($k$) to 
the flux variance. 

The cross spectrum, $P_\times (k_\parallel,\Delta \theta)$ is related to the
flux power spectrum by
\begin{equation}
P_\times (k_\parallel,\Delta \theta) = \int_{k_\parallel}^\infty 
\frac{d k^\prime}{2 \pi} k^\prime
J_0 \left[\sqrt{k^{\prime ^2} - k_\parallel ^2}\, u_\perp (\Delta \theta)\right] 
P_F (k^\prime, \frac{k_\parallel}{k^\prime})
\label{pcross}
\end{equation}
where $J_0$ is a zeroth order Bessel function. The above form follows from
azimuthal symmetry, the Bessel function coming from an integration over 
azimuthal angle. The dependence on cosmology is embedded primarily in the 
form of $u_\perp (\Delta \theta)$. 
$P_F (k^\prime, \frac{k_\parallel}{k^\prime})$ also depends on cosmology,
through redshift distortions (\S \ref{redshift distort}), but
the dependence should be weak for currently favored ranges of cosmologies.
Again we will generally plot the
dimensionless quantity, 
$k_\parallel P_\times (k_\parallel, \Delta \theta) /\pi$.
These definitions illustrate the relation between the auto and
cross spectra and the full three-dimensional flux power, but in practice
we measure the auto and cross spectra by one-dimensional 
Fast Fourier Transforms (FFTs) as described below.

\subsection{Simulation Method}
\label{sim_method}

We summarize here the model parameters used in simulating the forest.
The temperature of the IGM follows 
a power-law in the gas density, $T=T_0 (1 + \delta)^\alpha$, where
$\alpha$ should be between $0$ and $0.6$
(Hui \& Gnedin 1997). Here $\delta$
is the gas over-density, $\delta = (\rho -\bar{\rho})/\bar{\rho}$, with
$\rho$ the gas density and $\bar{\rho}$ its mean, and $T_0$ is the temperature
at mean density. 
From photo-ionization equilibrium, it then follows that the optical depth 
of a gas element is related to the gas density by 
$\tau=A (1 + \delta)^\nu$. The power law index $\nu$ is related to the
power law index of the temperature-density relation by $\nu = 2.0 - 0.7 \alpha$
and $A$ is a parameter related to the intensity of the ionizing background 
which is adjusted to match the observed mean transmission. 
The optical depth
is then shifted into redshift space, taking into account the effects of
peculiar velocities, and then convolved with a thermal broadening window. 
We thereby have a recipe for creating artificial Ly$\alpha$ spectra
from realizations of the density and velocity fields, which cosmological
simulations provide us with. 

In this paper we wish to examine many different models, and will thus
use N-body only simulations as opposed to hydrodynamic simulations which
are much more computationally expensive. Our methodology will follow that
of Zaldarriaga et al. (2001a,2003): 
we produce a grid of models with different 
points in the grid corresponding to different models of the IGM, compute
the auto and cross spectra for each model in the grid, and compare with
observations. For the most part we rely on $256^3$ particle, $20.0$ Mpc/h 
box size, particle-mesh (PM) simulations, run assuming an Standard-Cold-Dark-Matter
(SCDM; $\Omega_m = 1$) cosmology.
On the range of scales probed by the Ly$\alpha$ forest, the linear
power spectrum is effectively power law, so running multiple simulations in 
an SCDM cosmology suffices to probe a range of power spectrum amplitudes
and slopes.
As we discuss in more detail in Appendix A, running the SCDM simulations
to approximate Lambda cosmologies means that we neglect the dependence of
dynamics and redshift distortions on $\Omega_m$, but the dependence should
be small since $\Omega_m (z \sim 3) \sim 1$ for the Lambda cosmologies 
considered.
The details of the simulations are provided in Appendix A,
where we also show resolution and boxsize tests.

Our N-body simulation supplies us with a realization
of the dark matter density and velocity fields.\footnote{We use 
TSC interpolation to interpolate particle positions and velocities onto a 
mesh. We use the same number of mesh points as particles.}  
To obtain the hydrogen
gas density field and velocity field we smooth the dark matter density
and momentum fields in order to incorporate roughly the effects of gas 
pressure. The smoothing is applied in k-space, after which the resulting
smoothed real space density field is obtained by a 
three-dimensional FFT. The smoothing
is described  in k-space by 
$\delta_g(k) = exp \left( - \frac{k^2}{k^2_f}\right) 
\delta_{\rm dm} (k)$, where $\delta_g$ is the gas density, 
$\delta_{\rm dm}$ is the dark matter density, and $k_f$ describes the smoothing
scale. This simplified prescription for computing the gas density fields
seems likely to produce reasonably good agreement with fully hydrodynamic 
simulations, but more tests are 
warranted (Meiksin \& White 2001, Gnedin \& Hui 1998,McDonald 2003). 

Given the gas density and velocity fields, we generate mock Ly$\alpha$ 
spectra; the gas density is mapped into an optical depth, which is shifted
into redshift space, and convolved with a thermal broadening window. To summarize, then, our
model of the IGM thus has several free parameters:
$a$, the output scale factor of the simulation which is 
related to the power spectrum
normalization; $n$, the slope of the primordial power spectrum which
is related to the shape of the power spectrum on scales probed
by Ly$\alpha$ forest measurements; $k_f$, the
wave number of the pressure smoothing filter; $T_0$\footnote{We express 
$T_0$ in
units of (km/s)$^2$. The relationship between $T_0$ in units
of (km/s)$^2$ and $T_0$ in units of $K$ is 
$T_0 = \frac{T_0[(km/s)^2]}{165} 10^4 K$.}, the temperature of the
IGM at mean density; $\alpha$, the logarithmic slope of the temperature-density
relation; and $\langle f \rangle$, the mean flux in the Ly$\alpha$ forest.
The parameter, $A$, related to the strength of the ionizing background, is
adjusted to match the mean flux, $\langle f \rangle$, at each grid 
point in our grid of IGM models. To generate the cross spectrum we assume 
a flat universe and introduce the additional parameter $\Omega_\Lambda$.

To measure the auto spectrum from a simulation we 
generate mock Ly$\alpha$ spectra along 6,000 random
lines of sight, with the line of sight direction taken along each of the
different box axes, and measure the auto spectrum by one-dimensional FFT,
finally averaging the results over the different lines of sight.
For examining our fiducial model introduced below, 
in order to reduce sample variance errors, we average over several
different realizations of the density and velocity fields, each generated from
the same cosmology, but with differing initial phases. 

We then measure the cross spectrum as follows:
1) The desired angular separation between two lines of sight is converted
into  units of transverse cells by $u_\perp (\Delta \theta) =
\frac{H(z)}{H_{\rm box}(z)} D_A(z) \Delta \theta \frac{N^{1/3}}{L_{\rm box}}$ 
cells, assuming a particular cosmology. Here $H(z)$ and $D_A(z)$ are the Hubble
parameter and the co-moving angular diameter distance in the cosmology 
in which we wish
to calculate the cross spectrum and $H_{\rm box}(z)$ is the Hubble parameter 
in the cosmology of the simulation box; in this case an EDS cosmology. The
parameters $L_{\rm box}$ and $N$ are, respectively, the size of the 
simulation box
in units of Mpc/h and the number of grid points. 
2)  We generate a flux 
field for 3,000 random pairs of lines of sight, each separated by a transverse 
distance of $n_a$ cells, where $n_a$ is the closest integer {\it less} 
than the 
desired $u_\perp$ in cell units.
3) The same is done for an additional 3,000 random pairs of 
lines of sight, each of these pairs
separated by $n_b$ cells, where $n_b$ is the closest integer {\it larger} than
the desired $u_\perp$ in cell units. 4) We measure the cross spectrum from
the lines of sight separated by $n_a$ cells and from the lines of sight
separated by $n_b$ cells, and linearly interpolate between the two 
measurements to find the cross spectrum at the exact desired separation.
5) In considering the fiducial model introduced below, 
we average the results over five different realizations
of the density and velocity fields.

\subsection{Examples}

\begin{figure}[htb]
\centerline{\psfig{figure=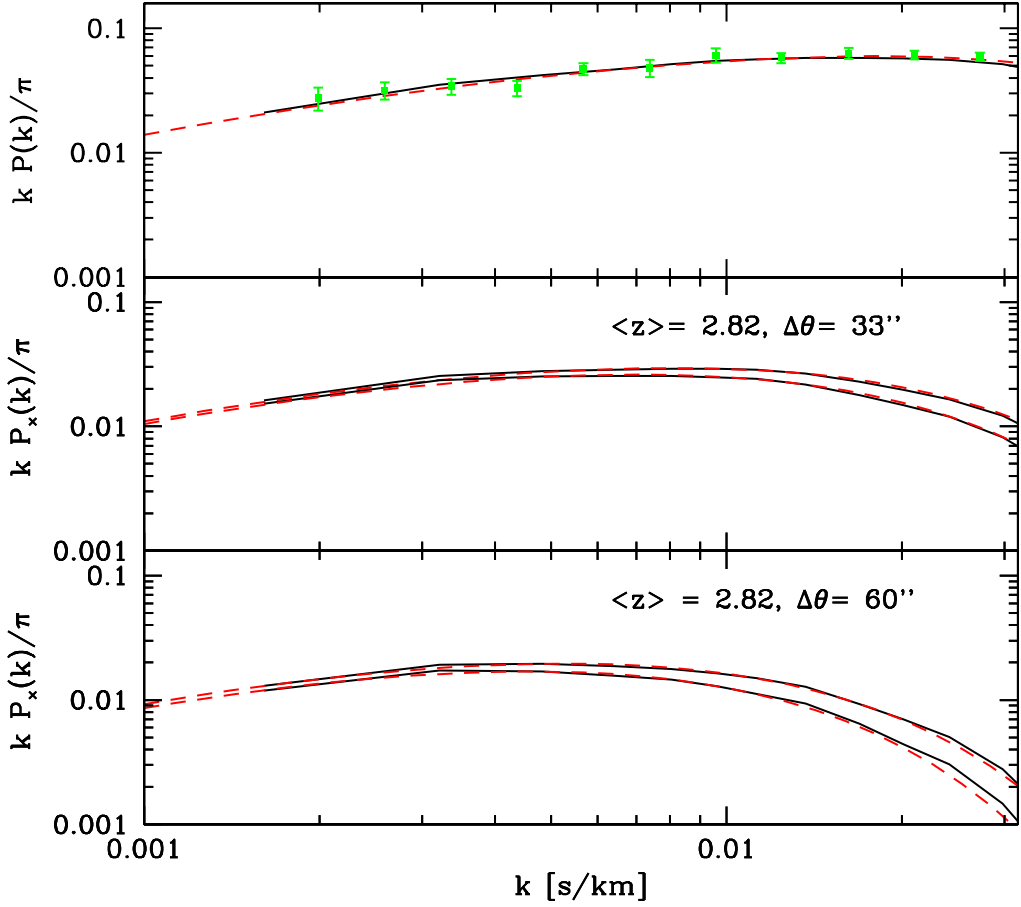,height=6.0in}} 
\caption{\label{mcdfit}
The top panel shows Croft's measured auto spectrum, interpolated to
$\langle z \rangle = 2.82$ (green points with error bars), 
a fit to this with the fitting
formula of \S \ref{predictions} (red dashed line), and a measurement from 
simulations (black solid line). 
The middle panel shows a fit to the cross spectrum
at separation $\Delta \theta = 33''$ (red dashed line), 
and the simulation measurement at the same separation (black solid line). 
The top black solid curve in the panel and its red dashed fit are for a
$\Omega_m=0.3$, $\Omega_\lambda=0.7$ (Lambda) cosmology, and the bottom curves
for a $\Omega_m=1.0$ flat (EDS) cosmology.
The bottom panel shows a fit to the cross spectrum at separation
$\Delta \theta = 60''$ (red dashed line) 
and the simulation measurement at the 
same separation (black solid line). 
Again the top curves in this panel are for a 
Lambda cosmology and the bottom curves for a EDS cosmology.
}
\end{figure}

From the auto and cross spectra measured in this way, we determine the
goodness of fit of a model in our parameter grid by calculating,
first for the auto spectrum, $\chi^2_{\rm auto} = \sum_{i=1}^n 
\left[\frac{\left(P_{\rm sim} (k_i)- P_{\rm obs} (k_i)\right)^2}{\sigma^2_i}\right]$. 
We then compute, as described in \S \ref{data}, the goodness of 
fit of the model
to the observed cross spectrum, using an analogous $\chi^2$. 
We add a term coming from observational
constraints on $\langle f \rangle$ to our calculated value of 
$\chi^2_{\rm auto}$. At $z=2.82$ this term is
$\chi^2 = \left[\left(\langle f \rangle - 0.682 \right)/0.034\right]^2$, 
where the value 0.682 is 
from Press at al.'s (1993) measurement of the mean transmission, which, at this
redshift, is in good agreement with recent measurements by 
Bernardi et al. (2002) from
the SDSS quasar sample, although recently some authors have argued that these
measurements may be biased low (Seljak, McDonald \& Makarov 2003). The
uncertainty in the mean transmission is taken to be $5\%$ following
Zaldarriaga et al. (2003).
In the equation for $\chi^2_{\rm auto}$ 
the sum runs over the data from Croft et al. (2002) with wavenumber
$k_i \lesssim 0.03$ s/km,
interpolated to $z=2.82$. Here $P_{\rm obs} (k_i)$ is the 
observed auto spectrum
at wavenumber $k_i$, $\sigma_i$ is Croft et al.'s (2002) observational error
estimate for $P_{\rm obs} (k_i)$, and $P_{\rm sim} (k_i)$ is 
the simulation measurement
interpolated to $k=k_i$. 
A resolution test shown in Appendix A demonstrates that our 
fiducial simulation, with $256^3$ particles in a $20.0$ Mpc/h box, gives
a reliable (compared with present observational errors)
flux auto spectrum on large scales, but misses some power in the
auto spectrum on small scales. 
Because of this, we do not include
the auto spectrum data at $k \gtrsim 0.03$ s/km in constructing our fits. 
(A plot of the auto spectrum is shown later in the text, in figure 
\ref{mcdfit}.)

From now on, whenever we refer to the fiducial model for the purpose of
illustration, it refers to the following:
$(a, n, k_f, T_0, \alpha, \langle f \rangle) = 
(0.19, 0.7, 35.0$ h$^{-1}$ Mpc, $250$ (km/s)$^2$, $0.2, 0.684)$.
Similar models were found to fit the observed auto spectrum by Zaldarriaga
et al. (2001a,2003). The power spectrum 
normalization and slope of this model can also be characterized 
by $\Delta^2 (k_p, z=2.82)$ and $n_{\rm eff}$, 
which are the amplitude and slope of the linear power 
spectrum at $k_p=0.03$ s/km, $z=2.82$ (Croft et al. 2002). 
This model 
has $\Delta^2 (k_p,z=2.82) = 0.44$, $n_{\rm eff} = -2.7$.
The fit has $\chi^2_{\rm auto} = 6.7$. 
The model has roughly 4 free parameters, $a, n, T_0$, and $\alpha$, 
since $k_f$ and
$\langle f \rangle$ are fixed, and we fit to 11 data points, so there
are 7 degrees of freedom. The fit, with $\chi^2$ per degree of freedom
of $0.96$ is reasonable. 

\begin{figure}[htb]
\centerline{\psfig{figure=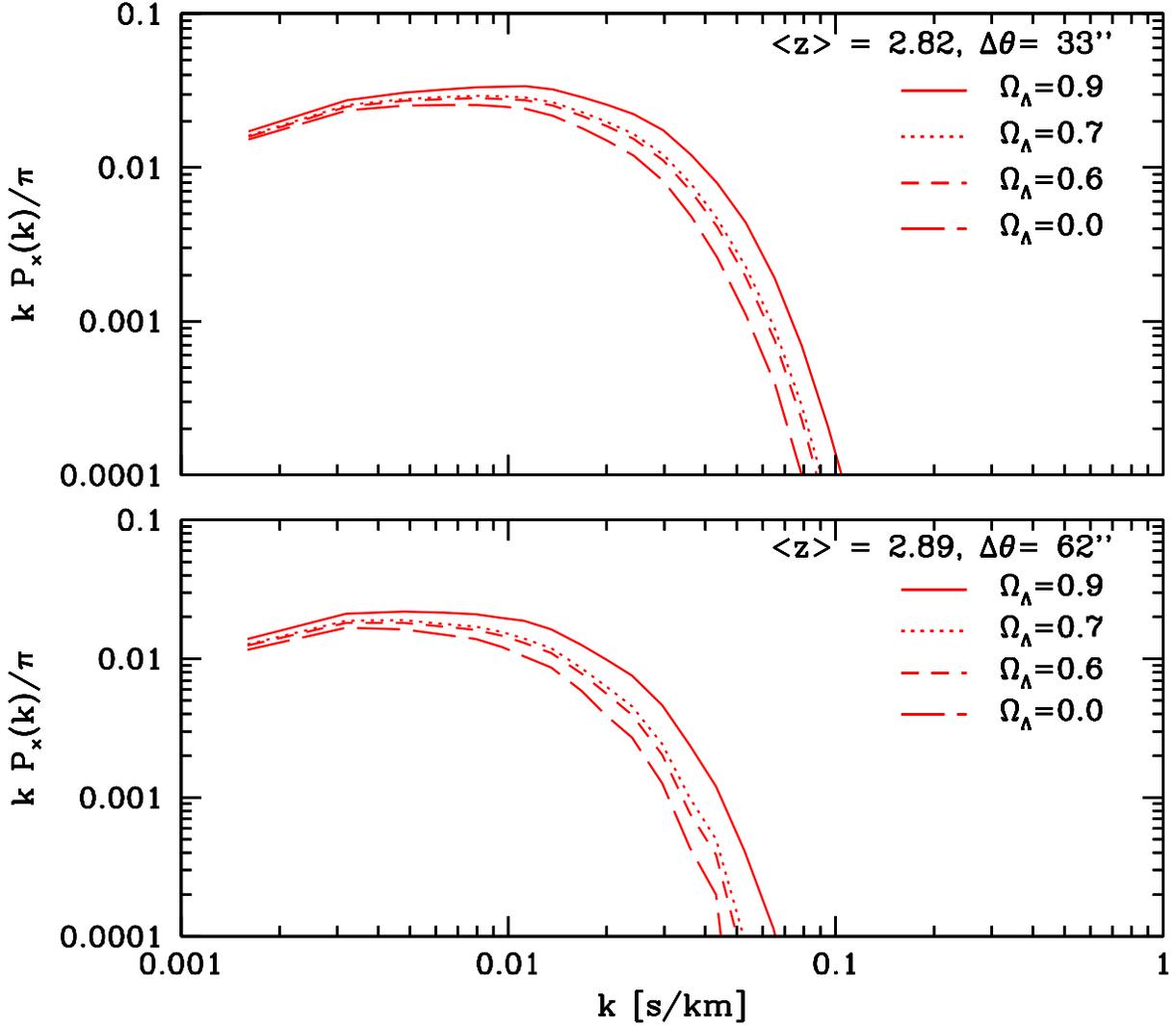,height=7.0in}}
\caption{\label{crossmods}
The cross spectrum for a range of flat cosmologies at 
$\langle z \rangle = 2.82, 2.89$,
assuming a model of the IGM that provides a reasonable fit to the
observed auto spectrum at each redshift. The top panel shows the cross
spectrum for a separation of $\Delta \theta = 33 ''$ and the bottom panel
shows the cross spectrum for a separation of $\Delta \theta = 62 ''$.
These redshifts and separations correspond to those in the data set considered
in \S \ref{data}.
}
\end{figure}

Since this fiducial model of the IGM provides a reasonable fit to the observed
auto spectrum it is interesting to examine the cross spectra this model
predicts for a range of cosmological geometries. 
From equation (\ref{fsquash}) we know how different the squashing
factors in different cosmologies are, but have yet to examine how
different the cross spectra are in these cosmologies. After all,
the cross spectrum is the observationally relevant quantity in our
study. 
We illustrate the difference in figure (\ref{crossmods}) 
for $\langle z \rangle=2.82, \Delta \theta = 33 ''$ and for
$\langle z \rangle=2.89, \Delta \theta = 62 ''$, which correspond
to the redshift and angular separation of two of the pairs we analyze in
\S \ref{data}.
The cross spectra shown in the figure are estimated from the average
of 5 different simulation realizations.
These predictions include the effect of redshift distortions as discussed
below in \S \ref{redshift distort}. 
Looking first at the $\Delta \theta = 33 ''$ plot, 
it will clearly be difficult to discriminate between a flat
$\Omega_\Lambda=0.7$ cosmology and a flat $\Omega_\Lambda=0.6$ cosmology,
as already implied by the similarity of the squashing factor for these
two cosmologies in figure (\ref{squash}). 
At $k_\parallel = 3.2 \times 10^{-3}$ s/km, which is twice the wavenumber
at the fundamental mode, the
cross spectrum in the $\Omega_\Lambda=0.7$ cosmology only differs from the
cross spectrum in an EDS cosmology by $\sim 7\%$, while the cross spectrum
in the $\Omega_\Lambda=0.6$ cosmology differs from the 
EDS case by only $\sim 5.5\%$. By $k_\parallel \sim 0.03$ s/km, the fractional 
differences with the EDS case are larger, around $\sim 33\%$ for 
$\Omega_\Lambda=0.7$ and $\sim 27\%$ for $\Omega_\Lambda=0.6$.   
However a cosmology with a
significant cosmological constant makes its presence felt, and there
is a sizeable difference between a flat, $\Omega_\Lambda=0.9$ cosmology and
a flat, $\Omega_\Lambda=0.7$ cosmology. The fractional difference between
the $\Omega_\Lambda=0.9$ case and the EDS case is $\sim 14\%$ at
$k_\parallel = 3.2 \times 10^{-3}$ s/km, and $\sim 53\%$ 
by $k_\parallel \sim 0.03$ s/km. Looking now at 
the $\Delta \theta = 62 ''$ plot,
the difference between models is larger, but the magnitude of 
the cross spectrum 
is reduced slightly and the scale at which the cross spectrum 
turns over moves to 
lower $k_\parallel$. At this separation, the fractional difference between
the $\Omega_\Lambda=0.9$ case and the EDS case is $\sim 21\%$ at
$k_\parallel = 3.2 \times 10^{-3}$ s/km and $\sim 73\%$ by 
$k_\parallel \sim 0.03$ s/km. 
As we describe in \S \ref{predictions}, however, the ability to 
discriminate between two cosmologies depends not on the fractional
difference between the cross spectra in the two models, but rather on
the difference between the cross spectra, divided by the square root
of the sum of the squares of the true auto and cross spectra.
This means that the discriminating power is actually better for the
close separation, $\Delta \theta \lesssim 1'$, pairs.
However, as we will flesh out in \S \ref{predictions}, 
the advantage provided by observing close separation pairs 
is nullified if the spectral resolution is not good
enough to resolve the small scales where such models differ significantly.
One can see some noise in the simulation measurement from 
the $\Delta \theta = 62''$ pairs at high $k$. Simulation measurements of the 
cross spectrum at large separations and high $k$ are generally rather noisy. 
This is not too much of a concern for our present purposes since 
in this paper we consider an observational data set of 
relatively close pairs, with poor spectral resolution, confining our 
measurements from data to large scales. 
In \S \ref{predictions}, however, we consider the expected cross spectrum
signal at large separations, extrapolating our simulation measurements to 
these separations using the fitting formula proposed by McDonald (2003).

\section{The Flux Power Spectrum and its Redshift Distortion}
\label{redshift distort}

In order to use the AP test to determine the cosmological constant, we must
take into account the effect of redshift distortions. It is our aim in this section to
investigate the relative importance of cosmology versus redshift distortions
in determining the anisotropy of the observed power spectrum. 
It is important to emphasize that while we will make comparisons between
simulations and linear theory predictions, we will not make use of the linear
predictions in the rest of the paper.

The detailed form
of the full three-dimensional flux power spectrum involves non-linear effects,
the effects of peculiar velocities and of thermal broadening. 
A start towards understanding these effects was made by Hui (1999), (see also 
McDonald \& Miralda-Escud\'e 1999), who made a
linear theory calculation of the effect of redshift distortions and 
found that $P_F(k,\mu) \propto (1 + \beta \mu^2)^2 e^{-\frac{k^2 \mu^2} 
{{k_s}^2}} P(k)$. (Further 
investigations have been carried out by McDonald 2003). 
Here $\mu$ refers to the ratio of the component of the wavenumber 
along the line of sight, $k_\parallel$, to the total wavenumber, $k$. The
optical depth is assumed to go as $\tau \propto (1+\delta)^\nu$ before
the effects of thermal broadening and redshift distortions are included;
$P(k)$ is the mass power spectrum in real space\footnote{The power spectrum
here is the baryonic power spectrum which is smoothed on small
scales relative to the dark matter, 
$P(k)=exp \left(-\frac{2 k^2}{k^2_f}\right) P_{\rm dm}(k)$}; 
and $\beta = f_g(\Omega_m,z) / \nu$. Here $f_g \equiv dlnD/dlna$ is 
the logarithmic derivative of the linear growth factor, $D(a)$, and 
$f_g(\Omega_m,z) \sim \Omega_m(z)^{0.6}$ for Lambda cosmologies. 
On large scales $P_F(k,\mu)$ has the same qualitative form as the expression
for the galaxy power spectrum in redshift space predicted by 
Kaiser (1987); the three-dimensional flux power is enhanced along the 
line of sight in redshift space. 
Here, however, the ``bias factor'' in the expression is 
given by $\nu$, the power law in the relation 
between density and optical depth.  
On small scales the flux power spectrum is suppressed along the line of sight,
as described by the factor $e^{-\frac{k^2 \mu^2} {{k_s}^2}}$. In linear
theory the smoothing scale, $k_s$, is the thermal broadening scale,
$k_s=\sqrt{2}/\sqrt{(T_0)}$, where $T_0$ is the temperature at mean
density in units of (km/s)$^2$. In the non-linear regime the flux power
spectrum will be suppressed on small scales not only from 
thermal broadening, but also from non-linear peculiar velocities and
the suppression will not generally be well described by a Gaussian.
This suppression of power is analogous to the finger of god effect 
found in the case of the galaxy power spectrum in redshift space, but here 
the effect is due not only to peculiar velocities but also
to thermal broadening. 
In this section we
quantify the anisotropy due to redshift distortions by direct measurement
from simulations.

We measure the three-dimensional flux power spectrum from a 
$512^3, 20.0$ Mpc/h simulation, assuming the fiducial model
described in the last section.\footnote{The 
temperature, $T_0= 300$ (km/s)$^2$, and the
temperature-density relation, $\alpha=0.1$, are slightly different than
the fiducial model from the $256^3$ simulation which has $T_0=250$ (km/s)$^2$
and $\alpha=0.2$. The model with the higher temperature, from the $512^3$ 
simulation, gives a slightly better fit to the observed auto spectrum, when 
all data points from Croft et al.'s (2002) 
measurement are included in the fit. See Appendix A for resolution tests.} 
To measure the full three-dimensional flux power spectrum, we form 
the flux field at every
pixel in the simulation box taking the line of sight along one of the box
axes, form the power spectrum by three-dimensional FFT, then repeat the
measurement with the lines of sight along the other box axes, and average the
results over the three box axes. 

\begin{figure}[htb]
\centerline{\psfig{figure=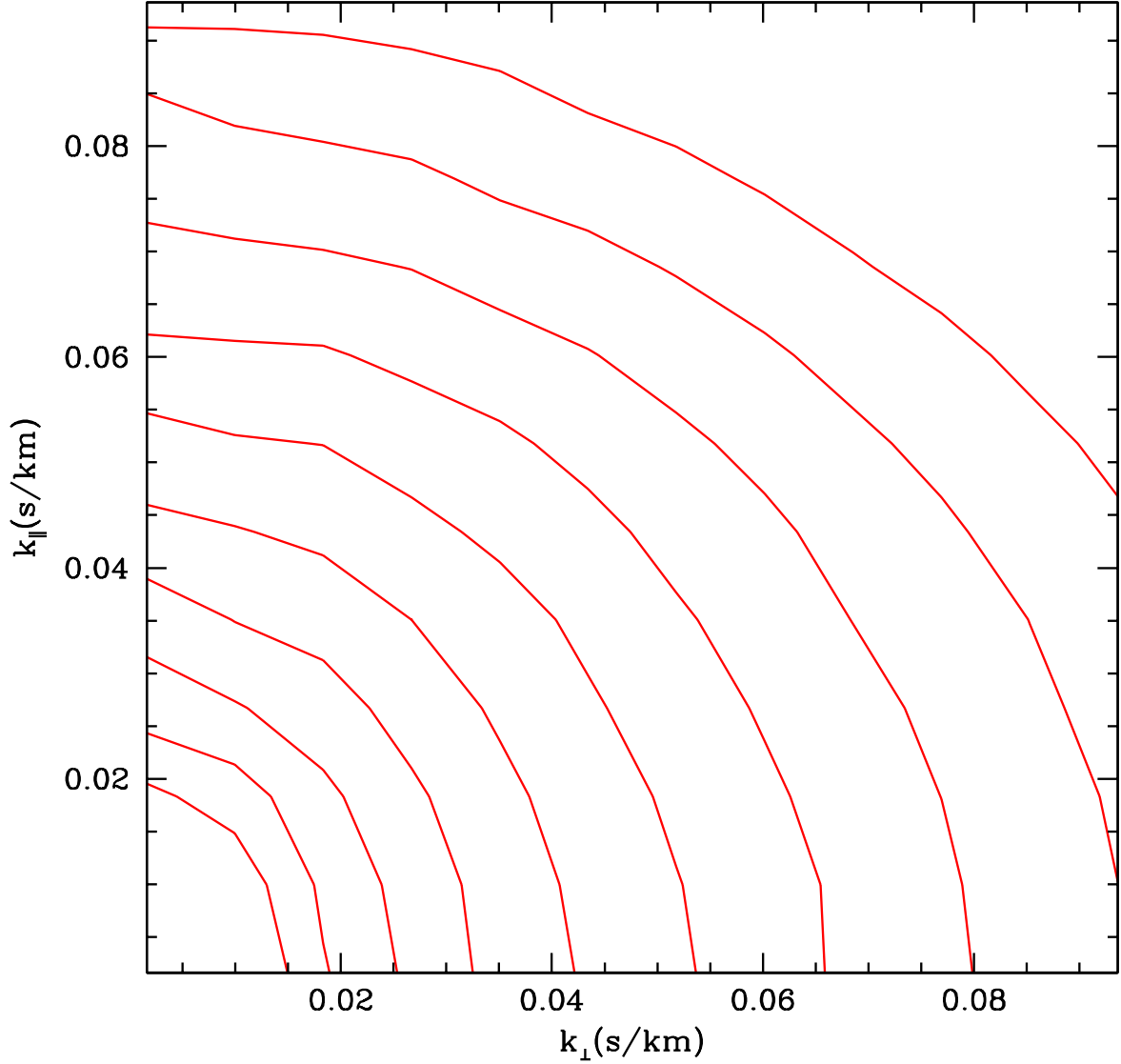,height=7.0in}}
\caption{\label{powcon}
Contours of constant flux power in the $k_\perp - k_\parallel$ plane.
The contours range from ln(P/(km/s)$^3$) = 6.5 to ln(P/(km/s)$^3$)=12.5.
The flux power is measured from an SCDM simulation with $512^3$ particles
in a $20.0$ Mpc/h box. The flux generation parameters are 
$(a,n,k_f,T_0,\langle f \rangle,\alpha) = 
(0.19,0.7,35.0$ h Mpc$^{-1}$, $300$ (km/s)$^2$, $0.684,0.1)$
}
\end{figure}

Our measurement is illustrated in figure (\ref{powcon}) where we show contours
of constant flux power in the $k_\perp - k_\parallel$ plane. On large scales
(small $k$) the contours of constant flux power have a slightly prolate shape.
This means that in configuration space contours of constant 
correlation function would
have a slightly oblate shape, signifying a squashing along the line of sight.
This behavior, of the flux correlation function in redshift space, is 
qualitatively the same as the behavior of the galaxy correlation in redshift
space; both are squashed on large scales from gravitational infall.
At higher $k$, near $k \sim 0.05$ s/km, the situation reverses itself 
and the contours of constant flux
power become oblate, signifying a finger-of-god elongation along the line
of sight. This behavior is also exactly analogous to the behavior of the
galaxy redshift space power spectrum and is qualitatively consistent with
the predictions of Hui (1999), McDonald \& Miralda-Escud\'e (1999), and
McDonald (2003). 

\begin{figure}[htb]
\centerline{\psfig{figure=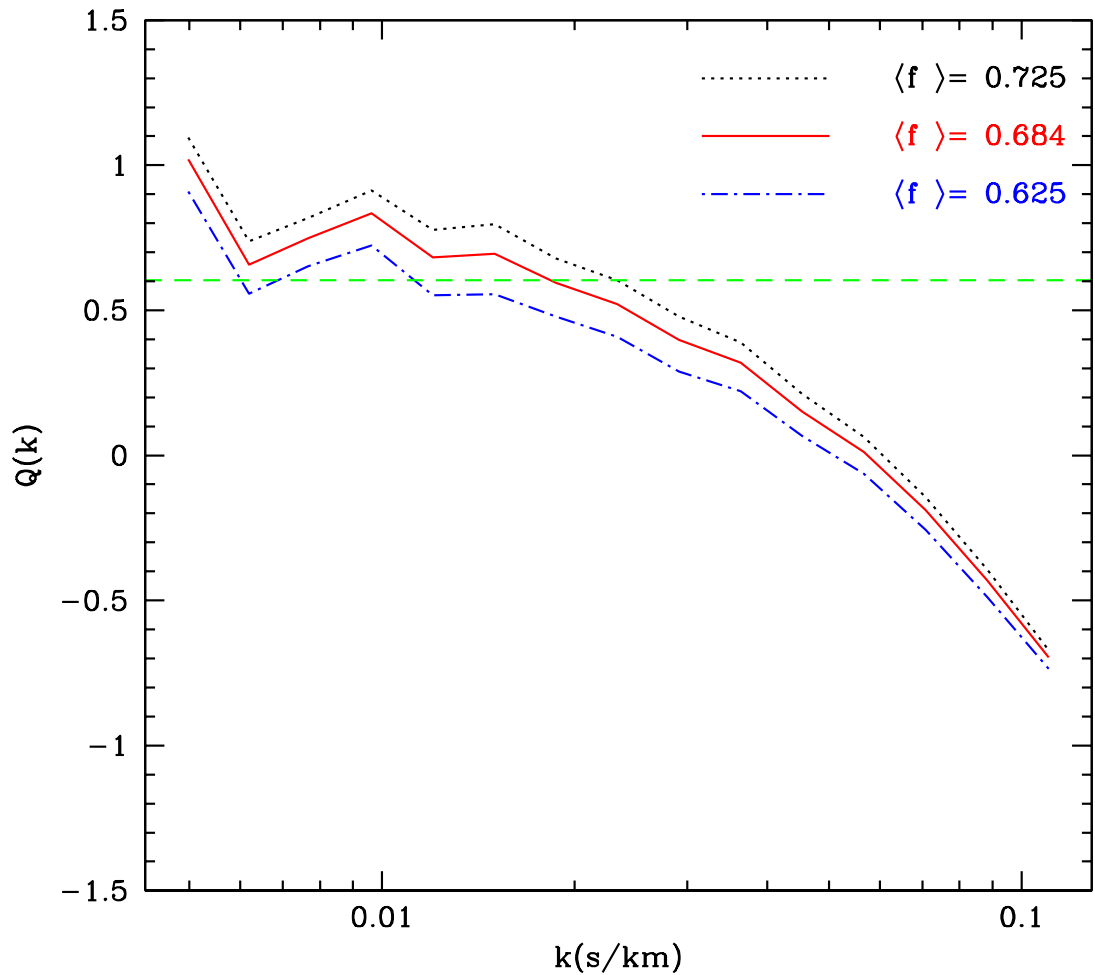,height=6.5in}}
\caption{\label{qtomeanf}
The quadrapole to monopole ratio of the flux power spectrum. The
red solid line is a simulation measurement with the same parameters as in
figure (\ref{powcon}). The simulation measurement is shown 
starting from $k = 3.0 \times k_{\rm fundamental}$.
The green dashed line is the large scale linear theory prediction,
which ignores the finger-of-god effect on small scales. 
We also show the quadrapole to monopole ratio for two models,
identical to the above model, except with differing values of 
the mean transmission, $\langle f \rangle$, demonstrating the sensitivity of
the measurement to $\langle f \rangle$.
Note that only the model with $\langle f \rangle = 0.684$ provides
a good fit to the auto spectrum.
}
\end{figure}

A useful statistic to quantify the measured anisotropy of the 
flux power spectrum is
its quadrapole to monopole ratio $Q(k)$, as is often done for the galaxy
power spectrum in redshift space. (See e.g., Cole, Fisher \& Weinberg 1994,
Hatton \& Cole 1998, Scoccimarro, Couchman \& Frieman 1999.) 
To compute this quantity one writes 
the anisotropic flux power spectrum, $P_F(k,\mu)$, as a sum of 
Legendre polynomials. The Legendre sum is 
$P_F(k, \mu) = \sum_{l=0}^{\infty} a_l (k) L_l (\mu)$, where $L_l (\mu)$
are the Legendre polynomials, and we wish to determine the coefficients,
$a_l(k)$, in the expansion. 
Using the orthogonality of the Legendre polynomials
we then have the quadrapole to monopole ratio,
\begin{equation}
Q(k) = \frac{a_2 (k)}{a_0 (k)} = \frac{5/2 \int_{-1}^1 d\mu {1 \over 2} 
\left(3 \mu^2 - 1 \right)
P_F(k,\mu)}{1/2 \int_{-1}^1 d\mu P_F(k,\mu)}
\label{quadtomon}
\end{equation}
We have measured the quadrapole to monopole ratio at $z=2.82$ 
assuming the same IGM model parameters as above. 
A plot of the measurement is shown
as the red solid line in figure (\ref{qtomeanf}). 
The measurement is shown starting from three times the fundamental 
mode, $k= 3.0 \times k_{\rm fun}$, since on larger scales there are few
$k$ modes with which to estimate the anisotropy. 
The measurement is generally noisy
since it comes from only one simulation realization. 
The measured quadrapole to monopole ratio has the approximate form 
expected given Hui's (1999) linear theory calculation. 
From the linear theory form
of the flux power spectrum, one expects that on scales where thermal
broadening is negligible, $Q(k) = \frac {4 \beta /3 +
4 \beta^2 /7} {1 + 2 \beta /3 + \beta^2 /5}$. 
At small $k$, the quadrapole to monopole measurement from the simulation 
is positive and has approximately
the linear theory magnitude, but it appears to be slightly larger on the 
largest scales measured. 
On smaller scales, the quadrapole
to monopole goes negative due to the suppression of power along the line
of sight from the combined effects of peculiar velocities and thermal 
broadening.

We caution, however, that the quadrapole to monopole ratio is somewhat
sensitive to $\langle f \rangle$, which in turn is sensitive to small
scale physics, some of which may be missing from our relatively low resolution,
N-body simulations. 
It is not obvious that the linear theory calculation gives the correct
biasing factor, as Hui (1999) points out. As Hui (1999) emphasizes,
the redshift distortions of the flux power spectrum are complicated,
involving several transformations; a transformation from gas density to
neutral hydrogen density, a transformation of the optical depth into
redshift space involving both a shift by peculiar velocity and convolution 
with a thermal broadening window, and finally an exponentiation
to go from optical depth to flux. The effect then of small scale, non-linear 
behavior on the redshift distortions at large scales is not clear. We know,
at any rate, that small scale physics is important in determining $A$ 
(the parameter adjusted to match the observed mean transmission discussed
in \S \ref{sim_method}), which in turn affects the large scale flux power.
It is perhaps not terribly
surprising then that the quadrapole to monopole depends somewhat
on $\langle f \rangle$, since this parameter sets which range of mass 
overdensities the flux field, $\delta_f$, is sensitive to. 
The black dotted and blue dot-dashed lines in figure (\ref{qtomeanf}) show the 
quadrapole to monopole ratios for models with
$\langle f \rangle = 0.625,0.725$, illustrating the dependence on 
$\langle f \rangle$.\footnote{We caution however that 
all of the other parameters of the IGM remained fixed in this comparison, 
so that the models with high and low $\langle f \rangle$ do not provide a good
fit to the observed auto spectrum. These values of $\langle f \rangle$ are 
also disfavored by observations. These values of $\langle f \rangle$ suffice
to show, however, that the
large scale redshift distortions appear sensitive to small scale physics.}  
The quadrapole to monopole ratio
for the model with the largest $\langle f \rangle$ is a bit larger than the 
fiducial model while the model with the smallest $\langle f \rangle$ has a
smaller quadrapole to monopole ratio than the fiducial model. This means 
that the quadrapole to
monopole ratio is likely to have a different magnitude at different redshifts,
where the mean transmission is different. Indeed the fitting formula that McDonald (2003) found (see \S \ref{predictions}) for the three dimensional flux power spectrum at 
$\langle z \rangle \sim 2.25$ implies a somewhat larger quadrapole to monopole
ratio on large scales, which is qualitatively consistent with our expectations
given the larger mean transmission at that redshift.
After performing these measurements and making the caveat about 
the dependence of our results on $\langle f \rangle$, 
we leave a more systematic investigation of flux power redshift 
distortions for future work.

\begin{figure}[htb]
\centerline{\psfig{figure=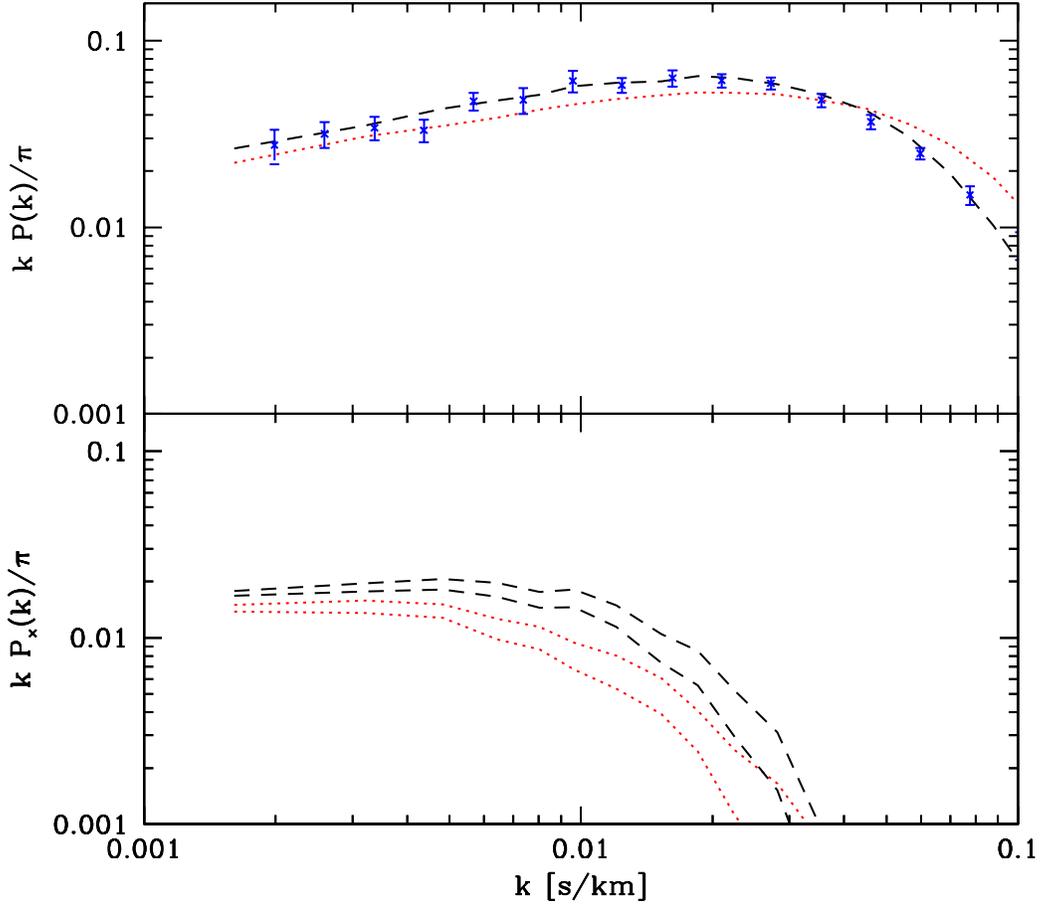,height=6.0in}}
\caption{\label{rdisto_v_cosmo}
The top panel shows (black dashed line) the auto spectrum from 
the $512^3$ simulation
including the effects of thermal broadening and peculiar velocities. The red 
dotted line is the auto spectrum in the same model, without the 
effect of peculiar 
velocities and thermal broadening. The blue points with error bars 
are Croft et al's (2002) measurement interpolated to 
$\langle z \rangle = 2.82$. 
The bottom panel shows (black dashed lines) the
cross spectrum measured at $\Delta \theta = 60''$ measured in a
Lambda model (top black dashed line) and in a EDS model 
(bottom black dashed line). The
red dotted lines are the cross spectra measured in the same models, ignoring
the effect of redshift distortions. {\it In calculating the cross
spectra in the redshift distortion free case, the mass power spectrum 
amplitude of the model has been adjusted so that the
corresponding auto spectrum matches observations on large scales.}  
}
\end{figure}

Given these measurements of the anisotropy of the three-dimensional flux power
spectrum due to redshift distortions, it is natural to ask how large the
anisotropy from redshift distortions is in comparison to the anisotropy
from cosmological distortions. To gauge the relative size of the two effects 
we consider the power spectrum of $f=e^{-A (1 + \delta)^\nu}$; the flux
power spectrum neglecting thermal broadening and peculiar velocities, which
is isotropic. We refer to this field as {\it the isotropic flux field} to 
distinguish it from the {\it anisotropic flux field} that includes redshift
distortions. In the top panel of figure (\ref{rdisto_v_cosmo}) we show a 
comparison of the auto spectrum measured from the fully anisotropic flux 
field and that measured from the isotropic flux field for
the same model of the IGM.\footnote{The isotropic and anisotropic models are
each normalized to have the same $\langle f \rangle$. The two models 
hence have different $A$s, where $A$ is the proportionality constant 
in the relation between optical depth and density.}  On large scales, 
the auto spectrum measured from
the anisotropic flux field is boosted compared to that measured from
the isotropic flux field. On the other hand, on small scales the auto 
spectrum measured from the anisotropic flux field is 
suppressed relative to that measured
from the isotropic flux field. This qualitative behavior of the auto
spectrum, that redshift distortions boost the auto spectrum on large scales 
and suppress the auto spectrum on small scales is expected from our
above measurements of the full three-dimensional flux power. How different
then our the cross spectra of the isotropic flux field and the anisotropic 
flux field? To provide a fair comparison, we first boost the mass power
spectrum normalization in the isotropic case so that the auto spectrum
of the isotropic flux field matches the observed auto spectrum on
large scales ($k \lesssim 0.03$ s/km). The  corresponding cross
spectra are shown in the bottom panel of figure (\ref{rdisto_v_cosmo}) 
for lines of sight separated by $\Delta \theta = 60''$ in each of a 
Lambda ($\Omega_\Lambda=0.7$) cosmology and an EDS cosmology. 
From the figure, one can see that redshift distortions significantly
boost the amplitude of the cross spectrum in each cosmology.
If one naively {\it neglects redshift distortions}, the cross spectrum one 
predicts in the Lambda cosmology
(top red dotted line) is less than that of the cross spectrum in the EDS 
cosmology {\it including redshift distortions} (bottom black dashed line)!
At larger separations, we expect that the geometrical effect will
become more important relative to the effect of redshift distortions. 
At any rate, this clearly suggests that redshift distortions are an
important effect and must be accounted for in detail in order to apply 
the Ly$\alpha$ AP test. Now that we have some sense of the 
size of the cosmological distortion and the
size of the redshift distortions, we measure the cross spectrum from a 
small data sample.

\section{Measurements of the Cross Power Spectrum from a sample
of Close Quasar Pairs}
\label{data}

In this section we present both a measurement of the cross spectrum
from our data, and present constraints obtained from comparison
with a small grid of simulated models. 
In Table (\ref{pairtab}) we provide a summary of pertinent information for the quasar pairs that we analyze. Observations were performed with 
the 4-meter/RC spectrograph combination at
both Kitt Peak National Observatory and Cerro Tololo Inter-american
Observatory, using the T2KB CCD detector and grating BL420 in 2nd order, and
the Loral 3K CCD and KPGL1 grating (1st order), respectively, delivering
3150\AA$ \la \lambda \la 4720$\AA\ and 3575\AA$ \la \lambda \la 6675$\AA\
for the KP triplet and 2139-44/45 pairs, respectively.
Further details of these (and redder observations for the KP triplet, plus
further spectra for additional objects) are detailed in Crotts \& Bechtold
(2003).

In order to estimate the cross spectrum of a given quasar pair the following
procedure is followed:

\begin{itemize}

\item We extract the region of the Ly$\alpha$ forest that is common to each
member of a quasar pair, avoiding regions close to each quasar that might
be contaminated by the proximity effect. (Bajtlik et al. 1988)  
We make a conservative cut for the proximity effect, including only
the region in the Ly$\alpha$ forest corresponding to rest frame
wavelengths of $\lambda_{\rm rest} \sim 1051$\AA $- 1185$\AA.
We then discard regions that do not overlap in wavelength between each member
of a given pair. In the case of the KP triplet, we discard regions that do
not overlap across the entire triplet.

We have not made any attempt to remove metal absorption lines in the 
Ly$\alpha$ forest. The effect of metal lines on the flux power spectrum is 
expected to be small, except on scales smaller than the resolution of 
our present data (McDonald et al. 2000).

\item We determine the flux field, 
$\delta_f=(f - \langle f \rangle)/\langle f \rangle$, for each member of a 
quasar pair. Here $f$
is the flux at a given pixel and $\langle f \rangle$ is the mean flux. To do
this we follow the procedure of Croft et al. (2002). We smooth the 
entire spectrum with a large radius, $50$ \AA, Gaussian filter giving 
the smoothed number of counts,
$f_s(\lambda)$, as a function of wavelength. From the smoothed spectrum, 
$\delta_f$ is given by 
$\delta_f=\left(f(\lambda) - f_s(\lambda)\right)/f_s(\lambda)$. 
This procedure 
estimates the mean flux directly, avoiding a measurement of the unabsorbed
continuum level. Clearly, however, this procedure
does not allow one to reliably measure the 
flux power on scales $r \gtrsim 50$ \AA, but
the flux power on these scales is expected to be contaminated by
continuum fluctuations anyway (Zaldarriaga et al., 2001b). We confine
ourselves to measurements of cross power on scales of $k_\parallel \gtrsim
2.0 \times 10^{-3}$ s/km.

\item We estimate the cross spectrum from the following
estimator, utilizing 1d FFTs:
\begin{equation}
\hat{P}_\times (k_\parallel ,\Delta \theta) = \frac{1}{n_k} 
\sum_{i=1}^{n_k} \frac{1}{2} \left[\delta_a (k_i) \delta^*_b (k_i) +
\delta_b (k_i) \delta^*_a (k_i)\right]
\label{pcross_est}
\end{equation}
Here the sum is over the $n_k$ modes in a given k-bin, $k_i$ is the
ith mode in the bin, and 
$k_\parallel$ is the average wave number in the bin.
Our convention is that modes with $\vert k_i \vert$ and $- \vert k_i \vert$
fall within the same k-bin and that $k_\parallel$ is always positive,
so that $\vert k_i \vert$ and $- \vert k_i \vert$ count as two modes
towards $n_k$.
The Fourier amplitude $\delta_a (k_i)$ comes from
the Fourier transform of the flux field of the first spectrum in the pair, 
$\delta_b (k_i)$ from the Fourier transform of the second spectrum
in the pair and $\delta^*_a$, $\delta^*_b$ are their complex conjugates.
The cross spectrum is thereby a real quantity that contains information about
the relative phase of the Fourier coefficients, $\delta_a (k_i)$, 
$\delta_b(k_i)$, as well as their amplitude. 

\item An estimate of the error on the cross spectrum measurements is
made assuming Gaussian statistics, a procedure which seems to 
be supported on relevant scales by analyses from a large number of spectra 
from the SDSS (Hui et al., in preparation). 
With this assumption the error on
the cross-spectrum estimate is (see Hui et al. 2001 for a derivation
of the auto-spectrum variance)
\begin{equation}
\sigma_{\hat{P}} ^2 (k_\parallel) = \frac{1}{n_k} 
\left[\left(P_\times(k_\parallel, \Delta
\theta)\right) ^2 + \left(P_{\rm 1d} (k_\parallel) + b^a (k_\parallel)\right) 
\left(P_{\rm 1d} + b^b (k_\parallel)\right)\right]
\label{err_est}
\end{equation}
Here $P_{\rm 1d}(k_\parallel)$ and $P_\times(k_\parallel, \Delta \theta)$ refer to the 
{\it true auto and cross spectra} as opposed to the noisy estimates from
the data, which we refer to as $\hat{P}_{\rm 1d}$ and $\hat{P}_\times$. In the
above equation $b^a (k_\parallel)$ and $b^b (k_\parallel)$ are 
estimates of the shot-noise
in each spectrum.\footnote{The shot-noise is estimated from the
quasar spectrum variance array as in Hui et al. (2001).} As in 
equation (\ref{pcross_est}), $k_\parallel$ refers to the average wavenumber
in a $k$-bin and $n_k$ is the number of modes in the bin counted with
the convention described previously.
When using these equations to estimate the error on the cross spectrum, 
we use simulation measurements to provide the true auto spectrum,
$P_{\rm 1d}(k_i)$ and the true cross spectrum, $P_\times(k_i, \Delta \theta)$.
The simulation measurements are made assuming a model for the IGM that fits
the observed auto spectrum at the relevant redshift. For the purposes of 
computing the error, we assume an $\Omega_m=0.3$, $\Omega_\Lambda = 0.7$ 
cosmology. For pairs with large separations, the cross spectrum contribution 
to the error is negligible. In practice we measure the cross power of
spectra that are smoothed by the limited spectral resolution of the
observations, and so we take 
$\sigma_{\hat{P}}(k_\parallel) \to \sigma_{\hat{P}}(k_\parallel) 
W(k_\parallel)$, where $W(k_\parallel)$ describes the smoothing due to
limited spectral resolution.

\item In deriving a constraint on $\Omega_\Lambda$ from 
the KP triplet, we need to take into
account that each pair is not independent of the other pairs. 
A given pairing of lines of sight from the triplet has a line of sight 
in common with every other pairing and so
measurements of the cross spectrum at a given scale are correlated
across the different pairs.
To take this correlation into account, we consider the
covariance between cross spectrum measurements from two pairs that
have a line of sight in common. Assuming Gaussian statistics, and
ignoring shot-noise which should be unimportant on the scales we 
probe reliably with the KP triplet, we find that the covariance is
\begin{equation}
\langle \delta P_\times^{12} (k_\parallel, \Delta \theta) 
\delta P_\times^{23} (k_\parallel, \Delta \theta^\prime) \rangle =
\frac{1}{n_k} \left[ P_\times^{12}(k_\parallel, \Delta \theta) P_\times^{23} 
(k_\parallel, \Delta \theta^\prime)
+ P_{1d} (k_\parallel) P_\times^{13} (k_\parallel, \Delta \theta^{\prime\prime}) \right]
\label{covar}
\end{equation}
The lines of sight are labeled by $1$, $2$, and $3$. We write
down the expression for the covariance between cross spectrum estimates
of pairs $12$ and $23$, so that line of sight $2$ is the common
line of sight to the two pairs, 
expressions for other pairings follow from swapping the super-scripts.
In this equation $P_\times^{12} (k_\parallel, \Delta \theta)$ is the
cross spectrum of lines of sight 1 and 2, which are separated by an
angle $\Delta \theta$, $P_\times^{23} (k_\parallel, \Delta \theta^{\prime})$ 
and
$P_\times^{13} (k_\parallel, \Delta \theta^{\prime\prime})$ have similar 
meanings, $P_{1d} (k_\parallel)$ is the auto spectrum, and $k_\parallel$,
$n_k$ have the meanings described above.
\end{itemize}

\begin{figure}[htb]
\centerline{\psfig{figure=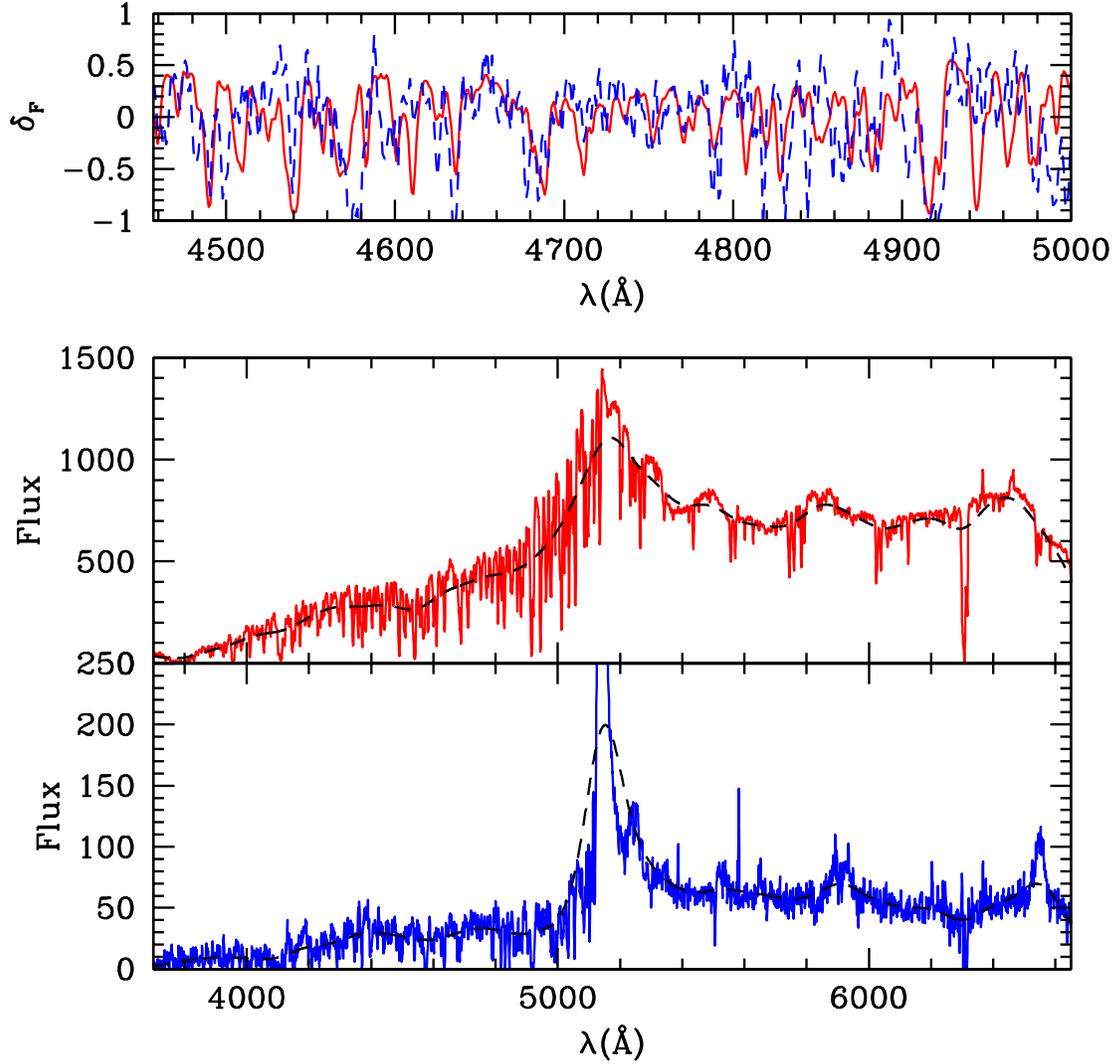,height=7.0in}}
\caption{\label{deltaform}
The bottom panel shows the spectrum of Q2139-4433 (red solid line) and
the middle panel shows the spectrum of Q2139-4434 (blue solid line). The dark
dashed lines show the spectra when smoothed with a Gaussian filter of 
$50 \AA$ radius in order to form $\delta_F$ (see text). The top panel 
shows $\delta_F$ overlayed for each spectrum. The red
solid line is $\delta_F$ for Q2139-4433 and the blue dashed line
is Q2139-4434. The pair is separated by $\Delta \theta = 62''$.
}
\end{figure}

In figure (\ref{deltaform}) we provide an example of our procedure for forming
$\delta_f$. The spectra of the members of the pair Q2139-33/34 are shown,
as well as the result of smoothing them with a $50 \AA$ filter. The flux 
fields, $\delta_f$, are shown one on top of the other, with 
the above cuts for the proximity effect. One can visually discern a 
cross-correlation between the two spectra in this pair, which has a separation
of $62 ''$. We generate flux fields for the other pairs in a similar fashion.

We then compare our measurements with a grid of models, using the method
of Zaldarriaga et al. (2001a,2003). 
In particular for each of the two high redshift
pairs we generate models with \\
$a=(0.11, 0.14, 0.19, 0.24)$, $n=(0.7, 0.8, 0.9, 1.0)$, 
$k_f=35.0$ h Mpc$^{-1}$,
$T_0 = (200, 250, 300)$ (km/s)$^2$,\\ $\alpha=(0.0, 0.2, 0.4, 0.6)$,
$\langle f \rangle = (0.625, 0.666, 0.700)$ at $\langle z \rangle = 2.89$ and
$\langle f \rangle = (0.650, 0.684, 0.725)$ at $\langle z \rangle = 2.82$, 
$\Omega_\Lambda =(0.0, 0.1, 0.2, 0.3, 0.4, 0.5, 0.6, 0.7, 0.8, 0.9, 1.0)$.\footnote{Our treatment of redshift distortions for the high
$\Omega_\Lambda$ models is probably inaccurate, since for these models
$f(z \sim 3) \sim \Omega_m(z \sim 3)^{0.6} \neq 1$. For instance,
if $\Omega_\Lambda =0.9$ then $f(z=2.26) = 0.87$. Given the size of our other 
uncertainties, we defer a more careful treatment to the future.
}   
At $\langle z \rangle =2.26$,
our grid spans a different range of power spectrum normalizations,
$a=(0.24, 0.27, 0.31)$ and mean transmissions, 
$\langle f \rangle = (0.750, 0.800, 0.850)$. In computing the likelihood of
each model in the grid, as described in \S \ref{sim_method}, we include
a prior constraint on $\langle f \rangle$. This constraint is
$\langle f \rangle = 0.666 \pm 0.033$ at $\langle z \rangle = 2.89$,
$\langle f \rangle = 0.682 \pm 0.034$ at $\langle z \rangle = 2.82$,
and $\langle f \rangle = 0.802 \pm 0.040$ at $\langle z \rangle = 2.26$. 
The values of $\langle f \rangle$ come from the measurements of 
Press et al. (1993).
In comparing the theoretical predictions with the observed auto spectrum,
since we compare with only one simulation realization, we do not include
points with $k \lesssim 2 \times k_{\rm fun}$ where $k_{\rm fun}$ is
the fundamental mode of the simulation box (see Appendix A). 
We do include such points
in our cross spectrum analysis because the observational error bars on
these points are very large and our results are not very sensitive to
whether we include these points or not.

For each model in our grid, we then compute $\chi^2_{\rm tot} =
\chi^2_{\rm auto} + \chi^2_{\rm cross}$. The computation
of $\chi^2_{\rm auto}$, the goodness of fit of the model auto spectrum
to the observed auto spectrum of Croft et al. (2002) interpolated to the
relevant redshift, is described in \S \ref{sim_method}.
To compute $\chi^2_{\rm cross}$ we take the sum $\chi^2_{\rm cross} =
\sum_{i=1}^n \left[\frac{(\widetilde{P}_{\rm \times, sim} (k_i, \Delta \theta) 
- P_{\rm \times, obs} (k_i,\Delta \theta))^2}{\sigma^2_i}\right]$. Here 
$P_{\rm \times, obs} (k_i,\Delta \theta)$ is the observed 
cross spectrum measurement
at wavenumber $k_i$ and transverse separation $\Delta \theta$, 
$\widetilde{P}_{\rm \times, sim} (k_i, \Delta \theta)$ is the 
simulation measurement
linearly interpolated to the same $k_i$ and $\Delta \theta$, and smoothed
to mimic the limited spectral resolution of our cross spectrum 
measurement.\footnote{For the KP triplet, we use the full covariance
matrix in calculating $\chi^2_{\rm cross}$, 
the off diagonal elements of which are calculated using equation
(\ref{covar}).}  
The error estimates of \ref{err_est} and \ref{covar} are multiplied
by the smoothing window before calculating $\chi^2$.
The spectral resolution of each spectrum is shown in
Table (\ref{pairtab}). We truncate the sum at $k_n$, the scale at which the 
limited spectral resolution causes the measured cross power to
drop to $50\%$ of the value it would have with perfect resolution.
From the vales of $\chi^2_{\rm tot}$ at each point in the parameter grid
we determine marginalized constraints on $\Omega_\Lambda$ following the
method of Tegmark \& Zaldarriaga (2000), Zaldarriaga et al. (2001a,2003). 

In brief, the method of Tegmark \& Zaldarriaga (2000) is as follows.
We begin from a grid of $\chi^2$ points in a seven-dimensional parameter
grid. We wish to find constraints on $\Omega_\Lambda$, marginalized
over all of the other ``nuisance'' parameters in the grid. To do this one
should integrate over the ``nuisance'' parameters. In this paper, we avoid 
the full multidimensional integration by assuming that the likelihood function
is multi-variate Gaussian, in which case integrating out a 
parameter is equivalent to 
maximizing the likelihood over the parameter (Tegmark \& Zaldarriaga 2000).
We then maximize over each parameter in turn, first cubic spline-interpolating
over the parameter. In this way we obtain marginalized constraints on 
$\Omega_\Lambda$.

\begin{figure}[htb]
\centerline{\psfig{figure=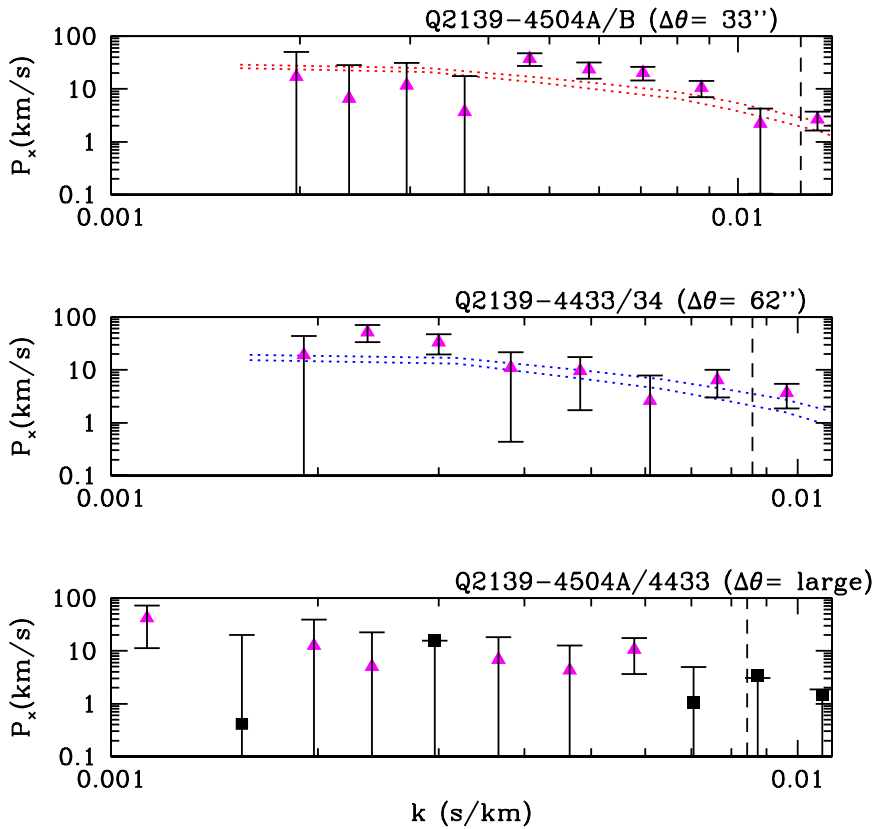,height=5.5in}}
\caption{\label{highz_pairs}
The top panel shows the cross spectrum measured from the pair Q2139-4504A/B.
We also show models from our grid that are close to the 
best fit, (to the combined auto and cross spectra), flat universe models 
with $\Omega_\Lambda=0.9$ (top curve) and $\Omega_\Lambda=0.0$ (bottom curve).
(Since $\chi^2$ is spline interpolated between grid points, the minimum
value we obtain for $\chi^2$ does not occur precisely on a grid point
in the parameter space. The curves shown are thus at grid points close
to this true minimum.)
The model fits have been multiplied by a filter representing the limited 
spectral resolution of the observations. 
The dashed line indicates the cut we make on the data due to the limited 
spectral resolution; only points to the left of the dashed line are 
included in our fit. The other IGM parameters are
the same for both models, $(a, n, k_f, T_0, \langle f \rangle, \alpha) = 
(0.19, 0.8, 35, 200, 0.684, 0.4)$. The models in the top panel are somewhat
poor fits to the observed cross spectrum, which has more power on 
small scales. 
The middle panel shows the same for the pair
Q2139-4433/34. In this case the other IGM parameters are 
$(a, n, k_f, T_0, \langle f \rangle, \alpha) = (0.14, 0.9, 35, 250, 0.666, 0.0)$.
and the models provide reasonable fits to the measured cross spectrum.
The bottom panel shows the cross spectrum of a widely separated pair, which
is consistent with zero. The magenta diamonds
indicate points with a positive cross spectrum and the black squares
indicate the absolute value of points with a negative cross spectrum.
}
\end{figure}

The cross spectrum measurements from the two high redshift pairs are
shown in the top two panels of figure (\ref{highz_pairs}).
From the figure one sees that the signal is generally 
weakly detected on large scales where few modes are available. 
On these scales some of the measurements in different k bins are 
consistent with zero at the $1 \sigma$ level. 
A more significant signal is detected on small scales where more
modes are available. Before comparing these cross spectrum 
measurements with models from the simulation grid, it is useful to
measure the cross spectrum from the overlapping regions in the
Ly$\alpha$ forests of a widely separated quasar pair.
If our underlying assumption that the cross spectrum signal arises
from fluctuations in the underlying density field is correct, the cross
spectrum of widely separated pairs should be consistent with zero.
In the bottom panel of figure (\ref{highz_pairs}) we show a measurement of the
cross spectrum between the spectra of Q2139-4504A/4433. Over the
range of scales that we use here for our constraints, the cross spectrum 
of the widely separated pair is consistent with zero. Quantitatively, we 
find that $\chi^2$ for the measurement, compared with the null hypothesis of
$P_\times(k_\parallel,\Delta \theta=\infty)=0.0$, is $5.0$ for $7$ degrees
of freedom. 

We then compare the measured cross spectra with fits
from the simulation grid. For the pair Q2139-4433/34, at
separation $\Delta \theta = 62''$ and $\langle z \rangle = 2.89$,
we find that the cross spectrum measurements are fit reasonably well by models
from the simulation grid. In particular the minimum $\chi^2$ is
$\chi^2_{\rm tot} \sim 11.2$ obtained near $\Omega_\Lambda=0.9$.
This is a reasonable fit to the 16 data points considered, (9 auto 
spectrum points and 7 cross spectrum points), given that our model
has roughly 5 free parameters. Each of the auto and cross spectrum
are well fit by the model; a model at a nearby grid point in the 
parameter space has $\chi^2_{\rm auto} \sim 5.7$. To gauge
the difference between models with different $\Omega_\Lambda$ we show models
in figure (\ref{highz_pairs}) with $\Omega_\Lambda=0.9$ and 
$\Omega_\Lambda=0.0$ that are close to the interpolated best fit parameters.
We emphasize that $\chi^2$ is spline interpolated between grid points and so 
the minimum value of $\chi^2$ does not, in general, 
lie on a grid point in the parameter space.
We include the plots of the model cross spectra just to provide some 
visual comparison between the data and the model fits.
The cross spectrum measurement of the pair Q2139-4504A/B is not
as well fit by the models in the parameter grid. The best fit model
in the grid has $\chi^2_{\rm tot} = 15.8$ obtained at $\Omega_\Lambda=1.0$.
The fit is to 18 data points (9 auto spectrum points and 9 cross 
spectrum points), with roughly 5 free parameters.
The auto spectrum is very well fit for these models with nearby grid points
having $\chi^2_{\rm auto} \sim 3.5$. The poor $\chi^2_{\rm cross}$ for this
pair is somewhat surprising. We searched the spectra for contaminating
metal line absorbers and for artifacts, but found no effects in the wavelength
range we analyze. Another possible issue is that we need to estimate our
resolution window carefully, since we have included scales where the power
is suppressed by the resolution by $\lesssim 50 \%$. We made a 
careful estimate 
by looking at the widths of narrow metal lines and night sky lines.
For now we consider the poor fit for Q2139-4504A/B to be a curiosity, 
but point out that larger data samples in the future will allow us to test 
more carefully for systematic effects.
 
\begin{figure}[htb]
\centerline{\psfig{figure=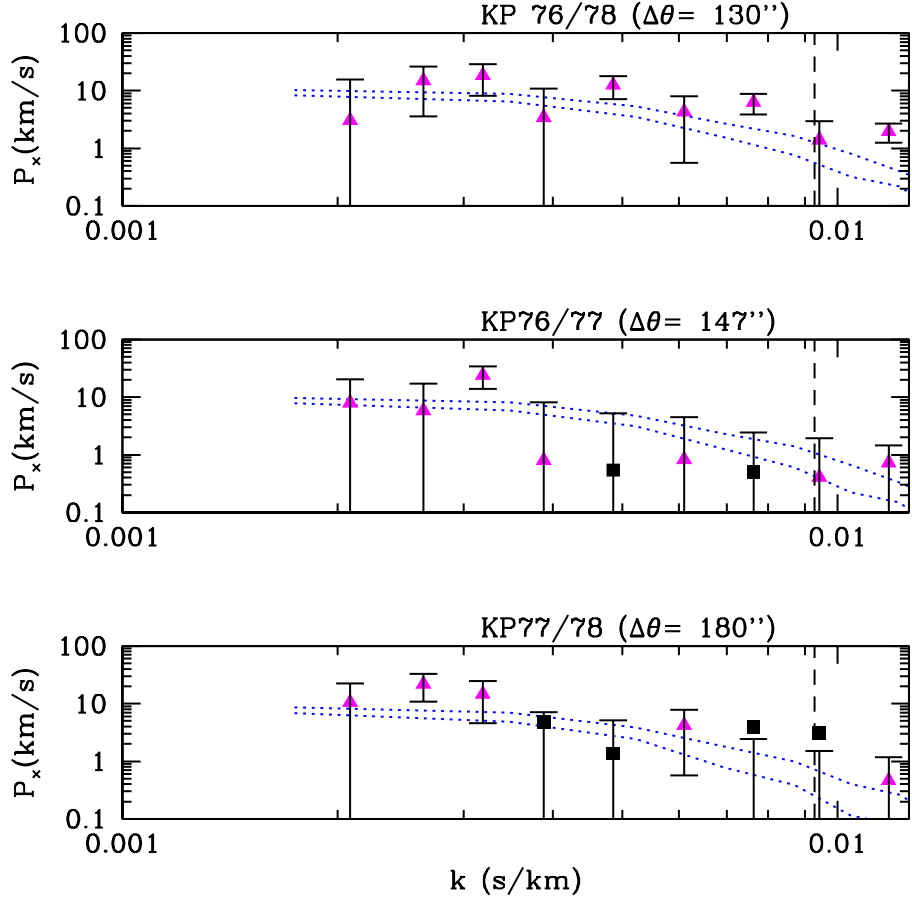,height=6.0in}}
\caption{\label{kptrip_pairs}
The panels show measurements of the cross spectrum from the KP triplet
pairs, as well as models from the simulation grid with 
$\Omega_\Lambda = 0.9$ and $\Omega_\Lambda=0.0$ to provide some
comparison between the data and model fits. 
The models have $(a, n, k_f, T_0, \langle f \rangle, \alpha) = 
(0.24, 0.9, 35, 250, 0.800, 0.6)$. The model fits have 
been multiplied by a filter representing the limited 
spectral resolution of the observations.
The magenta diamonds
indicate points with a positive cross spectrum and the black squares
indicate the absolute value of points with a negative cross spectrum.
In examining this figure, one should keep in mind that the measurements
of the cross spectrum at a given scale are correlated across the different
pairs, as discussed in the text.
}
\end{figure}

\begin{figure}[htb]
\centerline{\psfig{figure=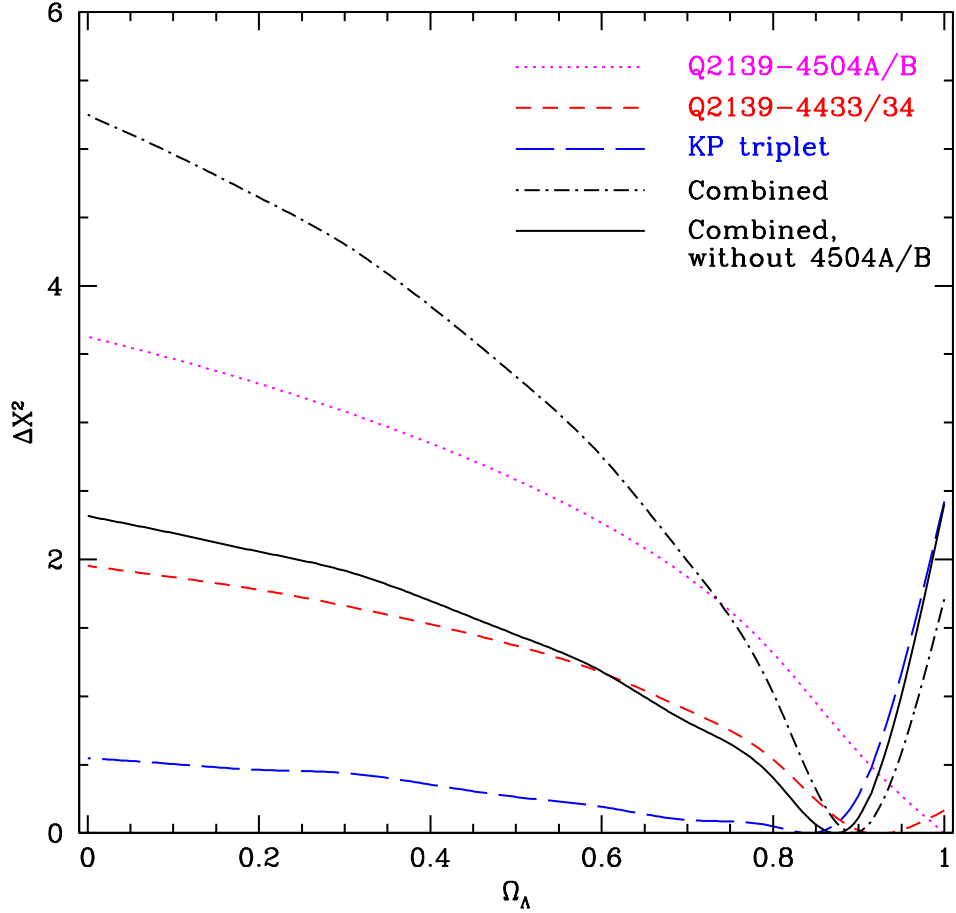,height=6.0in}}
\caption{\label{constraints}
The constraint on $\Omega_\Lambda$ from the quasar pairs.
The magenta dotted line shows $\Delta \chi^2 = \chi^2 - \chi^2_{\rm min}$
from the pair Q2139-4504A/B. The red short-dashed line shows the constraint
from the pair Q2139-4433 and the blue long-dashed line shows
the constraint from the KP triplet. The black dot-dashed line is the
combined constraint. The black solid line shows the constraint obtained
ignoring Q2139-4504A/B, since the cross spectrum of this pair is somewhat 
poorly fit by the models in our simulation grid.
}
\end{figure}

In figure (\ref{kptrip_pairs}) we show the measurements obtained from
the KP triplet. In this case the cross spectrum signal of the pairs,
with $\langle z \rangle \sim 2.26$ and separations in the range of
$\Delta \theta \sim 2' - 3'$, is only weakly detected. For the 
two larger separation pairs most of the
measurements of the cross spectrum at different scales are consistent with
zero. In spite of this, we still obtain a weak constraint on 
$\Omega_\Lambda$ from the KP triplet, taking into account the
covariance between the cross spectrum estimates of the three pairs. 
The best fit model is obtained
near $\Omega_\Lambda \sim 0.85$, where $\chi^2_{\rm tot} = 24.6$. This is
a reasonable fit given that there are 21 cross spectrum points and 9
auto spectrum points, and roughly 5 free parameters for a total of 25 degrees
of freedom.
We then combine the marginalized constraints on $\Omega_\Lambda$ by adding the
$\chi^2_{\rm tot}$ curve from the KP triplet with the constraints from
Q2139-4433/34 and Q2139-4504A/B.
Given that the cross 
spectrum of the pair Q2139-4504A/B is somewhat poorly fit by the 
models in the simulation grid, we
also consider the combined constraint ignoring this pair. Including 
Q2139-4504A/B,
the minimum total $\chi^2$ for the five pairs is $\chi^2_{\rm tot} \sim 52.5$,
(for $\sim 59$ degrees of freedom), obtained near $\Omega_\Lambda \sim 0.9$.
If we don't include the poorly fit pair, the total 
$\chi^2$ is $\chi^2_{\rm tot} \sim 36.1$ for the combined measurement, 
(for $\sim 41$ degrees of freedom) which occurs near 
$\Omega_\Lambda \sim 0.9$. 
The result, shown in figure (\ref{constraints}), is that,
if we include Q2139-4504A/B, $\Delta \chi^2 = \chi^2 - \chi^2_{\rm min}$
is $\sim 5.1$ for a flat model with $\Omega_\Lambda=0$, implying
such a cosmology is excluded at the $\sim 2.3 \sigma$ level.\footnote{When
we quote significances, we take a $2\sigma$ limit to be where 
$\Delta \chi^2=4$, or where the likelihood has fallen by a factor of $e^{-2}$
from its maximum value. Strictly speaking, this corresponds to a $95\%$ 
confidence limit only when the likelihood is
multivariate Gaussian, which is assumed in the way we perform the 
marginalization. (See Tegmark \& Zaldarriaga 2000).}
Ignoring the poorly fit pair,  $\Delta \chi^2$ 
reaches $\sim 2.3$ for $\Omega_\Lambda=0$. 
Our small sample of pairs seems to already disfavor 
an EDS cosmology at a significance of $\sim 1.5 \sigma$, or at an even larger
significance if one includes the constraint from the poorly fit pair.
There is a weak preference from each pair for a high value of 
$\Omega_\Lambda$; models with moderate $\Omega_\Lambda \sim 0.6$ are similarly disfavored at the $\gtrsim 1 \sigma$ level. Admittedly, even ignoring the
poorly fit pair, some our ability to rule out an EDS cosmology comes from
just two $k$-bins in the pair Q2139-33/34 with particularly large cross 
power (see figure \ref{highz_pairs}). While we have no reason to believe
the measurements in these $k$-bins are unreliable, we can check how 
much the significance depends on them. Ignoring
these two $k$-bins, we find that an EDS cosmology is still disfavored at the
1$\sigma$ level.

\section{Predictions for Future Samples/Observational Strategies}
\label{predictions}

Clearly it is not possible to draw any strong conclusions with the limited 
sample of five pairs analyzed above. This motivates considering the type of
sample necessary to do better in the future. In this section
we make predictions for the signal-to noise level (S/N) at which one can
distinguish between two different cosmologies with a given number
of quasar pairs, and examine the dependence of the expected S/N level
on observational resolution and shot noise. We hope that these considerations
will be useful in constructing an optimal observing strategy. This is similar
to estimates considered in (Hui, Stebbins \& Burles 1999), but here we 
use more accurate
expressions for the cross spectrum; our estimates coming from simulations
that include the effects of redshift distortions and the detailed form of
the three-dimensional flux power spectrum. These authors also
limited their considerations to large scales, ($k \lesssim 0.02$ s/km),
because of concerns about non-linear effects. We model the non-linear
effects with simulations and so can consider the benefit gained from
high $k$ modes, which can be quite substantial given close separation pairs and
high resolution data. In addition, we consider the
effects of instrumental resolution and shot-noise. McDonald (2003) has also 
made this type of comparison, primarily considering the expected
signal from pairs obtained with the SDSS. While our results are
broadly consistent with his, we emphasize the potential gain from
very close separation pairs observed with high spectral resolution.

The expected signal to noise level at which one can distinguish between
two cosmologies with the Ly$\alpha$ AP test is:
\begin{equation}
\frac{S}{N}
\sim \left[ \frac{f L}{\pi} \int_{k_\parallel = k_{\rm min}}^
{k_\parallel = k_{\rm max}} dk_\parallel \left[\frac{P^A_\times (k_\parallel, \Delta \theta) - P^B_\times (k_\parallel, \Delta \theta)} 
{\sqrt{\left(P_{\rm 1d}(k_\parallel) +b(k_\parallel)\right)^2 +
\left(P^A_\times(k_\parallel, \Delta \theta)\right)^2}}\right]^2 \right]^{1/2}  
\label{ston}
\end{equation}
In this equation $P^A_\times(k_\parallel, \Delta \theta)$ is the assumed true
cross spectrum, and $P^B_\times(k_\parallel, \Delta \theta)$ is the cross spectrum
in a model we would like to rule out.
The denominator is the Gaussian error estimate on the variance of the cross
spectrum given in equation (\ref{err_est}) for a single $k$-mode.\footnote{For
the shot-noise we assume $b(k_\parallel) \sim 1.0$ km/s. Below we consider the
effect of different assumptions about the shot-noise.} We caution 
that the Gaussian error estimate is likely to be inaccurate on small scales,
but leave a more careful treatment of errors for future work.
The integral is over all positive $k$-modes, from the lower limit of 
$k_{\rm min}$
to the upper limit of $k_{\rm max}$, with a density of states factor 
of $L/\pi$. 
For the length of the spectrum we take 
$L \sim 37,500$ km/s.\footnote{This is about $12,500$ km/s shorter than
the distance between Ly$\alpha$ and Ly$\beta$ at our fiducial redshift of 
$\langle z \rangle = 2.82$. This corresponds to a cut of about $25 \AA$ 
in the rest frame near each of Ly$\alpha$ and Ly$\beta$.} 
The factor $f$ is the fraction of 
the wavelength coverage between Ly$\alpha$ and Ly$\beta$ that is 
overlapping between the two spectra
in the pair, which we take to be $f=1$. The minimum wavenumber in 
the integral, $k_{\rm min}$ is taken to 
be $k_{\rm min} = 2.0 \times 10^{-3}$ s/km since larger scales are
expected to be contaminated by fluctuations in the continuum.
The maximum wavenumber in the sum is set by the spectral resolution
of the observation; it is taken to be the wavenumber
where the measured power falls to $1/e$ of its true vales.
Modeling the effect of spectral resolution as a Gaussian smoothing,
this wavenumber is given by $k_{\rm max} = \frac{1}{\sigma_s}$,
where $\sigma_s$ is related to the
instrumental FWHM by $\sigma_s = \frac{FWHM}{2 \sqrt{2 \ln(2)}}$.
In considering high spectral resolution data, we 
enforce an upper limit of $k_{\rm max} = 0.1$ s/km, since the
Ly$\alpha$ forest is expected to be contaminated by metal lines 
on such scales (McDonald et al. 2000).
One caveat is that in estimating our ability to discriminate between models in
this way, we assume perfect knowledge of the other parameters that go
into our modeling; the power spectrum normalization, the primordial
spectral index, the temperature of the IGM, etc. Measurements of the auto
spectrum from SDSS should dramatically reduce the error bars on the
observed auto spectrum, and make the associated errors in our modeling 
parameters smaller. 
Uncertainties in these modeling parameters will, however, affect our ability
to constrain $\Omega_\Lambda$, so the constraints quoted below are optimistic. 
We would like to estimate our ability to discriminate between cosmological 
models using pairs with a range of separations. Unfortunately, as described in
\S (\ref{sim_method}), direct measurements
of the cross spectrum at large separations from simulations are rather noisy. To get around
this difficulty, we use a fitting formula that describes the fully three-dimensional flux power spectrum and reproduces the results of simulation
measurements at small separations. This allows us to extrapolate cross
spectrum measurements to large angular separations. In particular, at our 
fiducial redshift
of $\langle z \rangle = 2.82$, we find the cross and auto spectra are
well fit if the three-dimensional flux power spectrum follows the 
fitting form found by McDonald (2003), $P_F(k, \mu) =
b_\delta^2 (1 + \beta \mu^2) D(k, \mu) P_L(k)$, with 
$D(k,\mu)=exp \left[\left(\frac{k}{k_{\rm NL}}\right)^{\alpha_{\rm NL}} - \left(\frac{k}{k_p}\right)^{\alpha_p} - \left(\frac{k_\parallel}{k_v (k)}\right)^{\alpha_v}\right]$ and 
$k_{\rm v}(k) = k_{\rm v0} (1+ k/k^{\prime}_v)^{\alpha_v ^ \prime}$. Here $P_L(k)$ is the linear power spectrum of mass and $D(k,\mu)$ describes
the effect of non-linear redshift distortions. The first term in $D(k,\mu)$ 
describes non-linear growth, the second term pressure smoothing, and the third
term describes the effect of non-linear peculiar velocities and thermal
broadening. The first two terms describe isotropic effects, while the third
term has a $\mu$ dependence since thermal broadening and peculiar velocities act only along the line of sight. Given this fitting formula we predict auto and 
cross spectra
by numerical integration using equations (\ref{pauto}) and (\ref{pcross}).
At our fiducial redshift of $\langle z \rangle = 2.82$ the 
cross and auto spectra are fit using the fitting parameters
$b_\delta^2 = 0.04$, $\beta=0.62$, $k_{\rm NL} = 10.0$ h Mpc$^{-1}$, 
$\alpha_{\rm NL} = 0.550$, $k_P = 14.0$ h Mpc$^{-1}$, $\alpha_p=2.12$,
$k_{\rm v0} = 5.50$ h Mpc$^{-1}$, $\alpha_v = 1.25$, $k^\prime _v = 0.917$ h Mpc$^{-1}$, and $\alpha^\prime _v = 0.528$. These parameters are different than
the ones found by McDonald (2003), but he considers a lower redshift, 
$z=2.25$, and uses a different linear power spectrum model. 
In figure (\ref{mcdfit}) we show the
fits to the auto and cross spectra from simulation measurements.
The
model fit to the auto spectrum is reasonable. The plot
also shows fits at separations of 
$\Delta \theta = 33''$ and $\Delta \theta = 60''$ for both EDS and 
Lambda cosmologies, which are reasonable. We also checked 
that the fitting formula provides
reasonable fits at $\Delta \theta = 90''$ in these cosmologies. 
With the
fitting formula in hand we then make predictions of the cross spectrum at
large separations using equation (\ref{pcross}). We then calculate our 
expected ability to discriminate between cosmologies using
equation (\ref{ston}).

\begin{figure}[htb]
\centerline{\psfig{figure=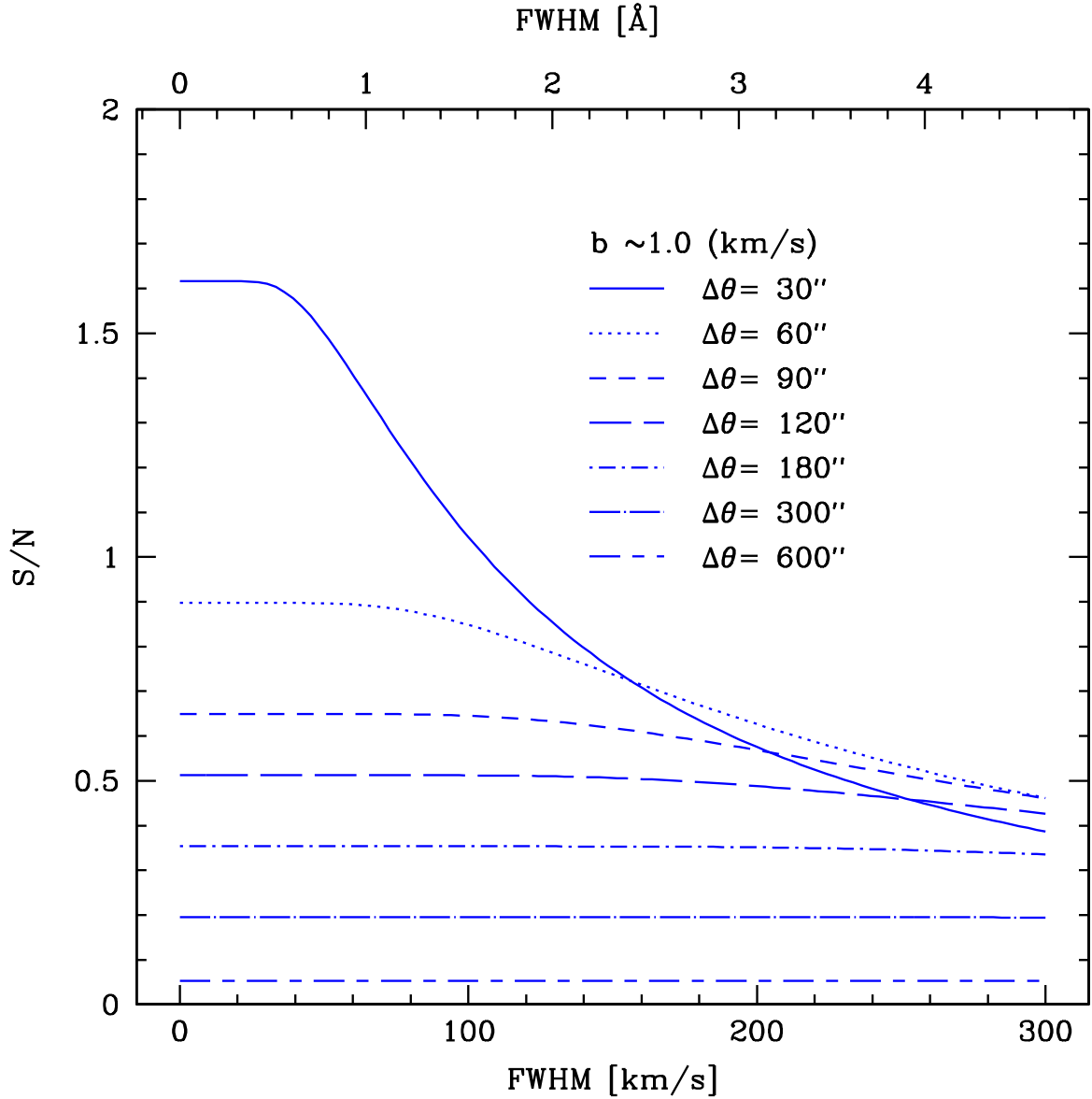,height=7.0in}} 
\caption{\label{ston_2.82}
The expected signal to noise level (S/N) at which one can distinguish
between an $\Omega_m=0.3, \Omega_\Lambda=0.7$ cosmology and a 
$\Omega_m=1.0$ cosmology from a single quasar pair as a function
of resolution (FWHM). The effect of shot noise is included, with
shot noise at the level of $b \sim 1.0$ km/s. We assume perfect knowledge
of all of the other parameters involved in modeling the IGM.
}
\end{figure}

In figure (\ref{ston_2.82}) we show the results of this calculation, in this 
case considering
the difference between an $\Omega_m=1.0$ (EDS) cosmology and a 
$\Omega_m=0.3, \Omega_\Lambda=0.7$ (Lambda) cosmology. For a 
single pair with $\Delta \theta = 30 ''$,
at high observational resolution, one can already distinguish 
between the two models
at $S/N \sim 1.6 \sigma$. Still considering high observational resolution,
we see that at larger separation, as the cross spectrum signal weakens, 
the S/N level goes down, reaching $S/N \sim 0.9 \sigma$ at $\Delta \theta = 60
''$ and $S/N \sim 0.35 \sigma$ at $\Delta \theta = 180 ''$. At still
larger separations, the discriminating power is quite weak, 
reaching $S/N \sim 0.2 \sigma$ at
$\Delta \theta = 300 ''$ and $S/N < 0.1$ at $\Delta \theta = 600 ''$.\footnote{For separations a bit smaller than $\Delta \theta = 30''$ the
discriminating power drops off.}

At lower observational resolution the advantage of measuring the cross spectrum
for close separation pairs is lost. When the FWHM becomes slightly worse 
than $\sim 150$ km/s, the S/N level, at $S/N \sim 0.7 \sigma$, 
is comparable for measuring the cross spectrum from a pair at 
separation $\Delta \theta = 30 ''$ and for using a pair at 
separation $\Delta \theta = 60 ''$.
In order to obtain the increased discriminating power of small separation
pairs, one must be able to resolve the high $k$ modes. At increased
angular separation, however,
observational resolution makes very little difference.
At these separations the cross spectrum (see eq. \ref{pcross}) turns over
at relatively low $k_\parallel$, and so, regardless of resolution, 
the high $k$ modes contribute
insignificantly to the S/N level. This explains why the S/N curves at large
angular separation in figure (\ref{ston_2.82}) are flat as a function of 
observational resolution. The conclusion is that {\it there is an advantage
to hunting out close separation pairs, but only if the observational
resolution is sufficiently good}. Below we give a rough formula to determine
what spectral resolution is good for observing pairs of a given separation.

\begin{figure}[htb]
\centerline{\psfig{figure=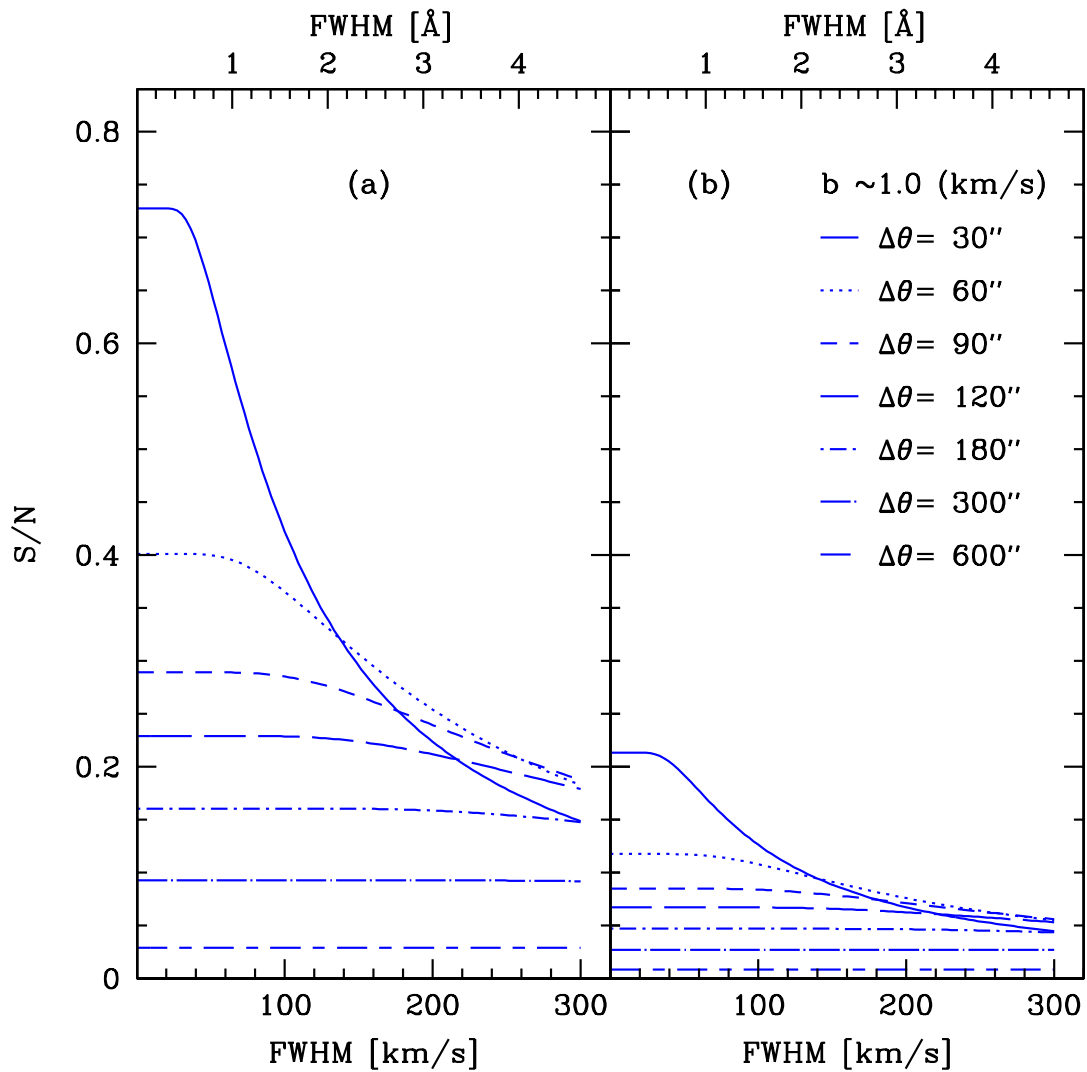,height=6.5in}}
\caption{\label{ston_oml}
a) The same as figure (\ref{ston_2.82}) except this plot shows the S/N level
at which one can distinguish between a $\Omega_m=0.3, \Omega_\Lambda=0.7$
cosmology and a $\Omega_m=0.2, \Omega_\Lambda=0.8$ cosmology from a single 
quasar pair.
b) The S/N level at which one can distinguish between two quintessence
cosmologies from a single quasar pair. Both cosmologies have $\Omega_m=0.3$
and $\Omega_Q=0.7$. The models have equations of state $w=-1$ and
$w=-0.7$. The equation of state is assumed constant with redshift.
In both a) and b), we assume perfect knowledge
of all of the other parameters involved in modeling the IGM.
It should be stressed that these comparisons are at $z \sim 3$.
At sufficiently lower or higher redshifts, the AP test should have an 
increased sensitivity to $w$.
}
\end{figure}

Figure (\ref{ston_2.82}) illustrates the possibilities of distinguishing between
an EDS cosmology and a Lambda cosmology, but it is more interesting to consider
the discriminating power between two more realistic cosmologies. In figure
(\ref{ston_oml}) we show the ability of the Ly$\alpha$ AP test to 
discriminate between a $\Omega_m=0.3, \Omega_\Lambda=0.7$
cosmology and a $\Omega_m=0.2, \Omega_\Lambda=0.8$ cosmology (left panel) as
well as the ability to discriminate between two quintessence cosmologies
(right panel).
At high observational resolution, using a single pair with separation
$\Delta \theta = 30 ''$, one can expect to 
distinguish between the two Lambda cosmologies at a level of $S/N \sim 0.7$. 
With a pair at $\Delta \theta = 60 ''$ the discriminating power falls
off to $S/N \sim 0.4$ and then to $S/N \lesssim 0.2$ at $\Delta \theta = 180
''$ and $S/N \lesssim 0.1$ at $\Delta \theta = 300 ''$. At the
moderate resolution of ${\rm FWHM} \sim 150$ km/s, where the discriminating
power is typically $S/N \sim 0.3$ for close separation pairs (with 
$\Delta \theta
\sim 30 - 120 ''$), one needs $\sim 200$ pairs to distinguish
between these models at the 4-$\sigma$ level. Alternatively, one
could achieve the same discriminating power with only $\sim 35$ close
separation ($\Delta \theta \sim 30''$), high resolution, pairs.
We can provide a rough criterion for determining the spectral resolution
that will be good for observing pairs at a given separation. We
adopt the rule that the spectral resolution, FWHM, should be
sufficiently good that the discriminating power drops off by no
more than a factor of $0.7$ from its value at optimal resolution.
Adopting this criterion for the case of discriminating between
the two Lambda cosmologies, {\it we find that the FWHM should be less than
than $80 (\Delta \theta/30'')^{1.1}$ km/s 
for pairs with separations between $\Delta \theta =30'' - 100''$ and 
that the spectral resolution is pretty irrelevant for larger separation
pairs. }
At separations between $\Delta \theta = 30'' - 600''$, we
find that the following fitting formula describes the expected 
discriminating power for a pair with a given separation and FWHM to
better than $10 \%$:
\begin{equation}
\frac{S}{N}(\theta,FWHM) \sim 
\frac{0.727 \left(\frac{30''}{\theta}\right)^{0.83} -
0.015 \left(\frac{300''}{\theta}\right)^{-1.1}}
{1 - \left(0.7 \ \frac{30''}{\theta} \frac{FWHM}{300 km/s}\right)
+ \left(3.0 \  \frac{30''}{\theta} \frac{FWHM}{300 km/s}\right)^2
- \left(1.6 \  \frac{30''}{\theta} \frac{FWHM}{300 km/s}\right)^3}
\label{ston_fit}
\end{equation}

In the right hand panel of figure (\ref{ston_2.82}), we show the ability of the Ly$\alpha$ AP test
to discriminate between a quintessence cosmology with equation
of state $w=-1$ and a cosmology with $w=-0.7$. Here the pressure of 
the quintessence
component is $p=w \rho_Q$, where $\rho_Q$ is the energy density of the
quintessence component, and we assume that the equation of state is
constant with redshift so that $\rho_Q = \Omega_Q (1+z)^{3 (1+w)}$. 
As mentioned in \S \ref{AP}, and illustrated in figure (\ref{squash}),
the Ly$\alpha$ forest AP test is barely sensitive to $w$, at $z \sim 3$. 
We will be making a small error here in the predicted signal for $w=-0.7$,
because the redshift distortions in this cosmology should be slightly
different than in the $w=-1$ cosmology. 
We find that, indeed, even with high observational resolution, the
discriminating power is only $S/N \sim 0.2$ for a single pair
separated by $\Delta \theta = 30 ''$. Discriminating between two quintessence
models with the Ly$\alpha$ forest AP test would require an extremely large 
number of close separation pairs. 
The Ly$\alpha$ forest AP test may, however, be 
helpful for attempts
to constrain quintessence models by helping to tighten constraints on 
$\Omega_m$, as mentioned in \S \ref{AP}, and emphasized by McDonald (2003).
Alternatively, applying the forest AP test to lower or higher redshifts
might be a more promising way of constraining $w$ (see figure \ref{squash}).

\begin{figure}[htb]
\centerline{\psfig{figure=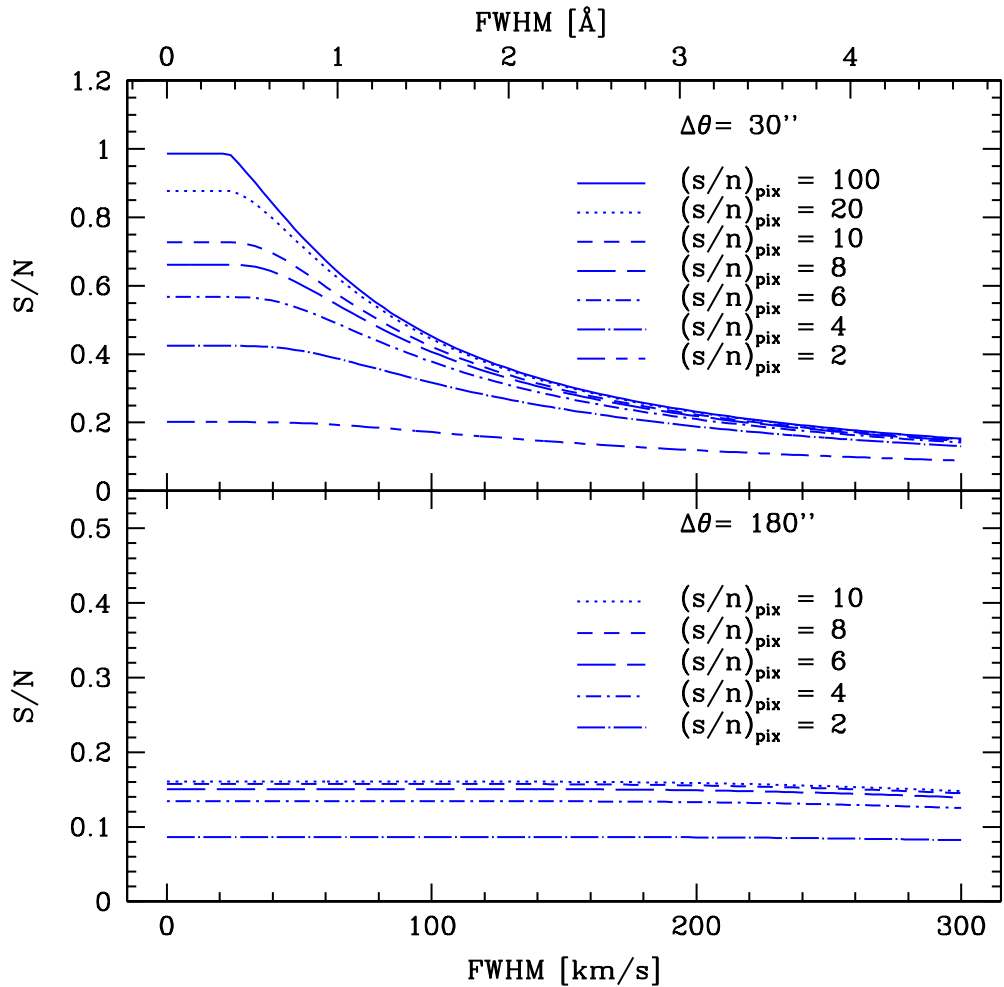,height=6.0in}}
\caption{\label{ston_noise}
The plot shows the effects of shot noise on the S/N level at which one
can distinguish between a flat $\Omega_\Lambda=0.7$ cosmology
and a flat $\Omega_\Lambda=0.8$ cosmology. 
The top panel is for a single quasar
pair separated by $\Delta \theta = 30.0 ''$ and the bottom panel
for $\Delta \theta = 180.0 ''$. The different curves represent
different assumptions about the typical signal to noise level per pixel,
$(s/n)_{\rm pix}$
in each quasar spectrum. In each case the signal to noise level
refers to the level at the continuum for a pixel size of
$\Delta u = 70$ km/s. The equivalent signal to noise level for a pixel
of arbitrary size, $\Delta u$, is 
$(s/n)_{\rm pix} \times \left(\frac{\Delta u}{70 km/s}\right)^{1/2}$.
We assume perfect knowledge
of all of the other parameters involved in modeling the IGM.
}
\end{figure}

It is also instructive to consider the effect of shot-noise on our ability
to discriminate between cosmologies. Since the noise fluctuations in each line of sight are independent, the noise has no cross-correlation and 
shot-noise does not directly contribute to
our estimator for the cross spectrum. However, shot noise does contribute 
to the variance of our estimate for the cross spectrum. Here we estimate the
shot noise level, at some typical signal to noise level per pixel, 
$(s/n)_{\rm pix}$, using an approximate
formula from Hui et al (2001). If the typical signal to 
noise level per pixel at the continuum is 
$(s/n)_{\rm pix}$, the pixel separation $\Delta u$, and the
mean transmission $\langle f \rangle$, then the shot-noise bias is
$b \sim \frac{\Delta u}{\langle f \rangle} \left((s/n)_{\rm pix}\right)^{-2}$.
For instance, in the above discussion we assumed $b \sim 1.0$ km/s, which for
$\langle f \rangle = 0.684$, $\Delta u = 70$ km/s ($\sim 1 \AA$ at 
$z \sim 2.82$), corresponds to a signal to noise per pixel 
of $(s/n)_{\rm pix} \sim 10$.
In figure (\ref{ston_noise}) we show the expected discriminating power (between
Lambda cosmologies with $\Omega_\Lambda=0.7$ and $0.8$)
for $(s/n)_{\rm pix} = 2, 4, 6, 8, 10, 20$ and $100$, 
with $\Delta u = 70$ km/s, 
for a close separation pair separated by $\Delta \theta = 30 ''$. 
We also show the discriminating power for a moderate separation pair,
with $\Delta \theta = 180 ''$, at shot noise levels corresponding to
$(s/n)_{\rm pix} = 2, 4, 6, 8, 10$.
Given the plot we can ask how low a shot-noise level is required so that the 
expected discriminating power, from a pair with a given
separation and spectral resolution, is degraded by no more than
a factor of $0.7$ from its optimal value. We will take this as a criterion
for determining what $(s/n)_{\rm pix}$ level is good for a pair at 
a given separation, observed with a given spectral resolution.
For the pair separated by $\Delta \theta = 30 ''$, at high spectral resolution,
the discriminating power is degraded by a factor of $0.7$ when
the signal to noise level reaches  $(s/n)_{\rm pix} \sim 10$.
At lower spectral resolution, FWHM $\sim 150$ km/s, the discriminating power 
is reduced by the same factor only when the signal to noise level reaches
$(s/n)_{\rm pix} \sim 3-4$. For the pair separated by 
$\Delta \theta = 180''$, the spectral resolution is unimportant and
the discriminating power is degraded by a factor of $0.7$ when
the signal to noise level reaches  $(s/n)_{\rm pix} \sim 3-4$.
In general shot-noise is an issue only when its contribution to the
cross spectrum error becomes comparable to the sample variance
contribution. The two contributions to the variance become comparable 
roughly when $b \sim P_{\rm 1d}(k)$; 
if $b \sim 1$ km/s, (corresponding to $(s/n)_{\rm pix} \sim 10$), 
the two are comparable near $k \sim 0.06$ s/km, at which scale the 
cross spectrum is very small unless the separation of the pair is also 
very small. With this level of shot-noise, the discriminating
power will only be degraded for high-resolution, close separation pairs. 
If the shot-noise is sufficiently bad, however, the
shot-noise contribution to the cross spectrum variance can become
comparable to the sample variance contribution even on large scales.
In this case, which occurs around $(s/n)_{\rm pix} \sim 3-4$, the
shot-noise will degrade the discriminating power even for large
separation pairs, as illustrated in the lower panel of 
figure (\ref{ston_noise}). {\it To summarize, for close separation pairs, with 
$\Delta \theta \sim 30''$, and high spectral resolution, given
a pixel size of $\Delta u$, a signal to noise level of
$(s/n)_{\rm pix} \sim 10 \left(\frac{\Delta u}{70 km/s}\right)^{1/2}$ 
is recommended. In other cases,
$(s/n)_{\rm pix} \sim 4 \left(\frac{\Delta u}{70 km/s}\right)^{1/2}$ 
is generally adequate.}

\section{Conclusions}
\label{conclusions}

In this paper we have described a concrete procedure for obtaining constraints
on dark energy from a version of the Ly$\alpha$ AP test.
We applied this procedure to a small sample of quasar pairs, and
discussed the possibilities of future observations. Our main results
are summarized as follows:

\begin{itemize}

\item While the Ly$\alpha$ AP test is a sensitive discriminator
of a cosmological constant, it is not sensitive to the equation of
state of a quintessence field unless $w(z \sim 3) \gtrsim -0.4$. 
(See also McDonald 2003.) The Ly$\alpha$ AP test is particularly
sensitive to the presence of a large $\Omega_\Lambda$.

\item The effect of redshift distortions on the cross spectrum are
important and must be taken into account in detail. 
At a redshift of $z=2.82$ we measure the quadrapole to monopole ratio
of the flux power spectrum and find that a linear theory prediction
is close to correct on large scales. The ratio depends somewhat
on $\langle f \rangle$. The lesson is that one needs simulations
to account for redshift distortions accurately.

\item We measured the cross spectrum from a sample of close quasar
pairs. The cross spectrum measurements have large error bars due to the
small number of pairs in the present sample. The measurements are generally 
consistent with the cross spectra from a grid of simulated models, if
perhaps weakly favoring high $\Omega_\Lambda$ cosmologies,
although one pair, Q2139-4504A/B, has somewhat excess small scale power.
A comparison between models from the simulation grid and the cross spectra
of Q2139-4433/34 and the KP triplet disfavors an EDS cosmology at a level
of $\sim 1.5 \sigma$. If we include the poorly fit pair, an EDS cosmology
is disfavored at $\gtrsim 2 \sigma$. Future, more accurate measurements of the cross spectra
will require comparison with a larger grid of simulated models.

\item We consider the expected power of future observations to discriminate
between different cosmological geometries.
A sample of $\sim 50$ moderate resolution, 
FWHM $\sim 150$ km/s, close separation pairs (with 
$\Delta \theta \sim 30''-120''$) should be able to discriminate between
an $\Omega_\Lambda=0.7$ and an $\Omega_\Lambda=0.8$ cosmology at the
$\sim 2 \sigma$ level, ignoring degeneracies with other IGM modeling
parameters. We find that there is a sizeable advantage obtained by
observing very close, $\Delta \theta \sim 30''$, separation pairs with
high spectral resolution, provided simulation measurements are reliable
at high $k$. Given $\sim 10$ high resolution, $\Delta \theta \sim 30''$ 
separation pairs, we should be able to distinguish between the two
Lambda cosmologies at the $2 \sigma$ level.

One should be able to obtain this type of sample with quasar pairs 
discovered by the Two Degree Field (2df) and SDSS surveys, provided
one does some follow-up spectroscopy (Burles, private communication 2003).
Two remarks may be helpful in this regard.
First, due to the finite optical fiber size, SDSS selects against quasar
pairs with separation less than 55''. One can attempt to recover pairs
with these separations by looking for objects with similar colors nearby
existing quasars. Second, SDSS has a spectroscopic cut-off at 3800 \AA.
One can recover many more pair spectra by going down to a lower cut-off of
$\sim 3400$ \AA. Since the auto spectrum has not yet been measured at
the low redshifts corresponding to these wavelengths, one would also
need to use the spectra in the sample to measure the auto 
spectrum at $z \sim 1.8$.

\end{itemize}

Finally, we mention a few words about future work. As the error bars
on cross spectrum measurements get smaller in the future, it will
be important to use higher resolution, larger volume simulations.
It will also be important to test the accuracy of our N-body plus
smoothing technique against full hydrodynamic simulations, especially if we
aim to compare data and theory at high $k$ where the hydrodynamic
effects should be most important. Measurements of the cross spectrum
may also be useful in other pursuits such as recovering the linear
mass power spectrum (Viel et al. 2002), reconstructing a three dimensional
map of the density field in the IGM (Pichon et al. 2001), and constraining
the effect of feedback on the IGM found by Adelbeger et al. (2003).

AL and LH are supported in part by the Outstanding Junior Investigator Award 
from the DOE, AST-0098437 grant from the NSF, and by
National Computational Science Alliance under grant AST030027, utilizing
the Origin 2000 array. AC acknowledges support from
grant AST-0098258 from the NSF and AR-9195 from STScI. MZ was supported 
by the David and Lucille Packard Foundation and the NSF. AL thanks the Fermilab
theoretical astrophysics department for hospitality. We thank Nick
Gendin for the use of his PM code, Roman Scoccimarro for discussions and
for reading a draft, and Scott Burles for discussions about
the expected number of quasar pairs from 2df and SDSS.

\section*{Appendix A}
\label{appendix}

In this section we mention some details regarding the simulations that
we run and show some convergence tests with respect to 
resolution and box size.

All of our simulations are run in an SCDM cosmology with the same linear
transfer function. We use the Ma (1996) transfer function, which is a modified
version of the BBKS transfer function (Bardeen et al. 1996). We assume a 
shape parameter of $\Gamma=\Omega_m h = 0.5$, and a 
normalization of $\sigma_8(z=0)=0.82$.
We run all simulations assuming an $\Omega_m=1$ cosmology so that the linear 
growth factor is just proportional to $1 +z$. In this case it is easy
to use different simulation outputs to represent different power spectrum 
normalizations. The box units are converted into units of km/s
assuming the EDS cosmology.\footnote{In converting the optical
depth to redshift space we take 
$s_\parallel = H_{\rm box}(z_{\rm sim}) x_\parallel /(1+z_{\rm sim}) + 
v_{\rm pec} (x)$ and 
$s_\perp = H_{\rm box}(z_{\rm sim}) x_\perp /(1+z_{\rm sim})$, where
$v_{\rm pec} (x)$ is the component of the peculiar velocity vector along
the line of sight.
In the event that we use a simulation output at $z_{\rm sim}$ to 
match the flux power spectrum at $z_{\rm obs}$, the redshift space 
coordinates are rescaled by a factor 
$[H_{\rm box}(z_{\rm obs})/H_{\rm box}(z_{\rm sim})] 
[(1 + z_{\rm sim}) / (1 + z_{\rm obs})]$.}
Although this is a rather disfavored shape for the initial linear power 
spectrum, our purpose here is to span a range of power spectrum
shapes and amplitudes near $z \sim 3$ {\it in units of km/s}, 
at scales around $1$ h/Mpc, 
for which this initial power spectrum template is adequate. 
In a different cosmology the conversion between units of
comoving $Mpc/h$ and observed $km/s$ units will require different initial
linear power spectra to fit the observed flux auto spectra; however the
differing initial power spectra must all lead to similar flux power spectra in
units of $km/s$ at $z \sim 3$. We do, however, then neglect the dependence
of dynamics and redshift distortions on $\Omega_m$. For flat models
with a cosmological constant, we should only be making a small error since
$\Omega_m(z \sim 3) \sim 1$. (For tests see Croft et al. 1998,
Gnedin \& Hamilton 2002). 
We examine a range of different power 
spectrum amplitudes and slopes. 
To investigate a range of power
spectrum normalizations, we save the density and velocity fields from the 
simulations at a range of output times. Different power spectrum shapes are
examined by considering a range of primordial spectral indices. 

\begin{figure}[htb]
\centerline{\psfig{figure=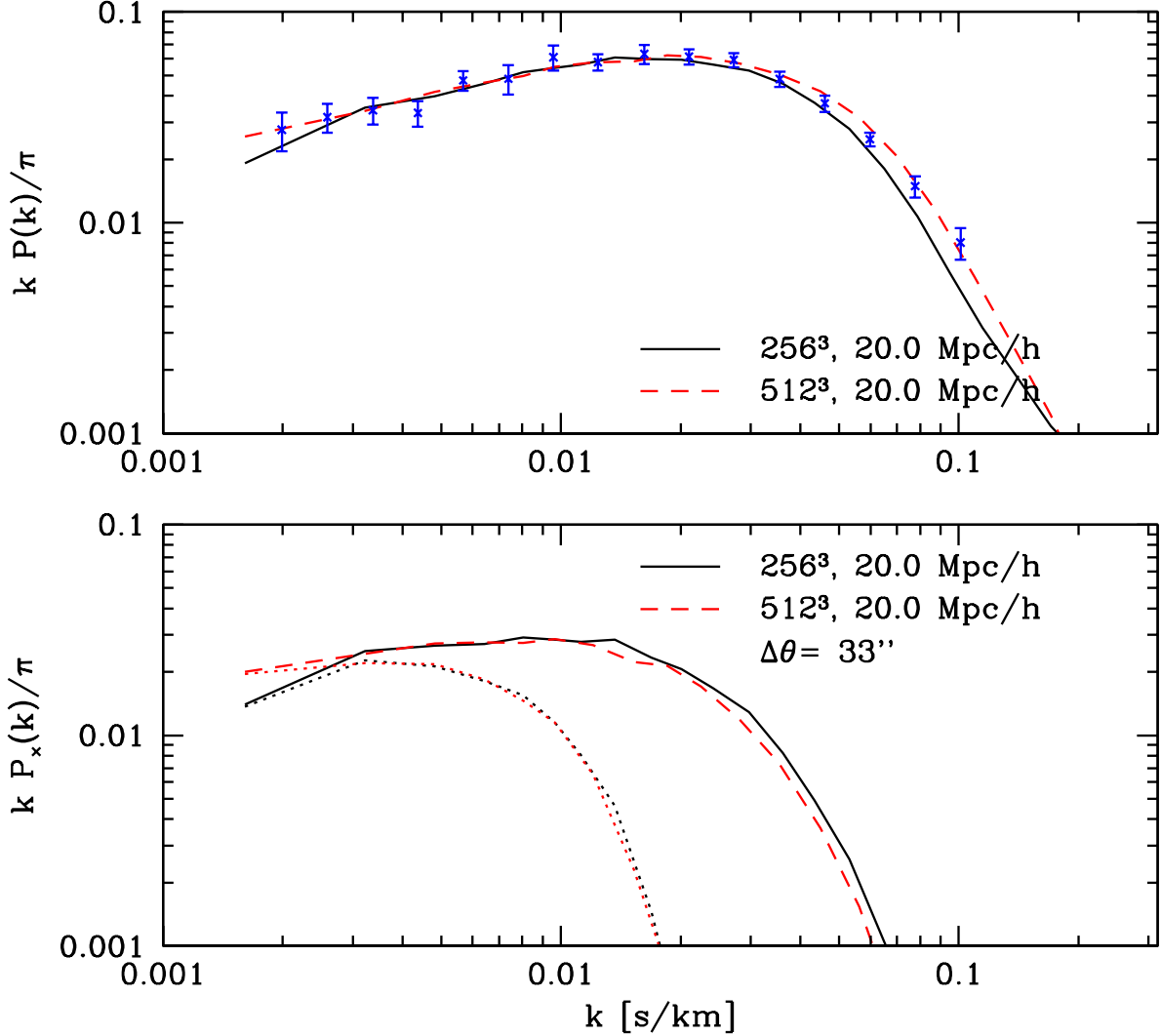,height=7.0in}}
\caption{\label{resolution}
A resolution test, showing the flux auto and cross spectra for a 
$256^3$ particle, $20.0$ Mpc/h boxsize simulation and those measured
from a $512^3$ particle, $20.0$ Mpc/h simulation in the same model.
The assumed IGM model is $(a, n, k_f, T_0, \alpha, \langle f \rangle) = 
(0.19, 0.7, 35.0$ h Mpc$^{-1}$, $250$ (km/s)$^2$, $0.2, 0.684)$. The cross 
spectrum is computed assuming $\Omega_m=0.3$. The comparison
is shown for a single simulation realization at each resolution. The
dotted curves show the cross spectra convolved with the instrumental
resolution of the close separation pairs studied in \S \ref{data}.
}
\end{figure}

\begin{figure}[htb]
\centerline{\psfig{figure=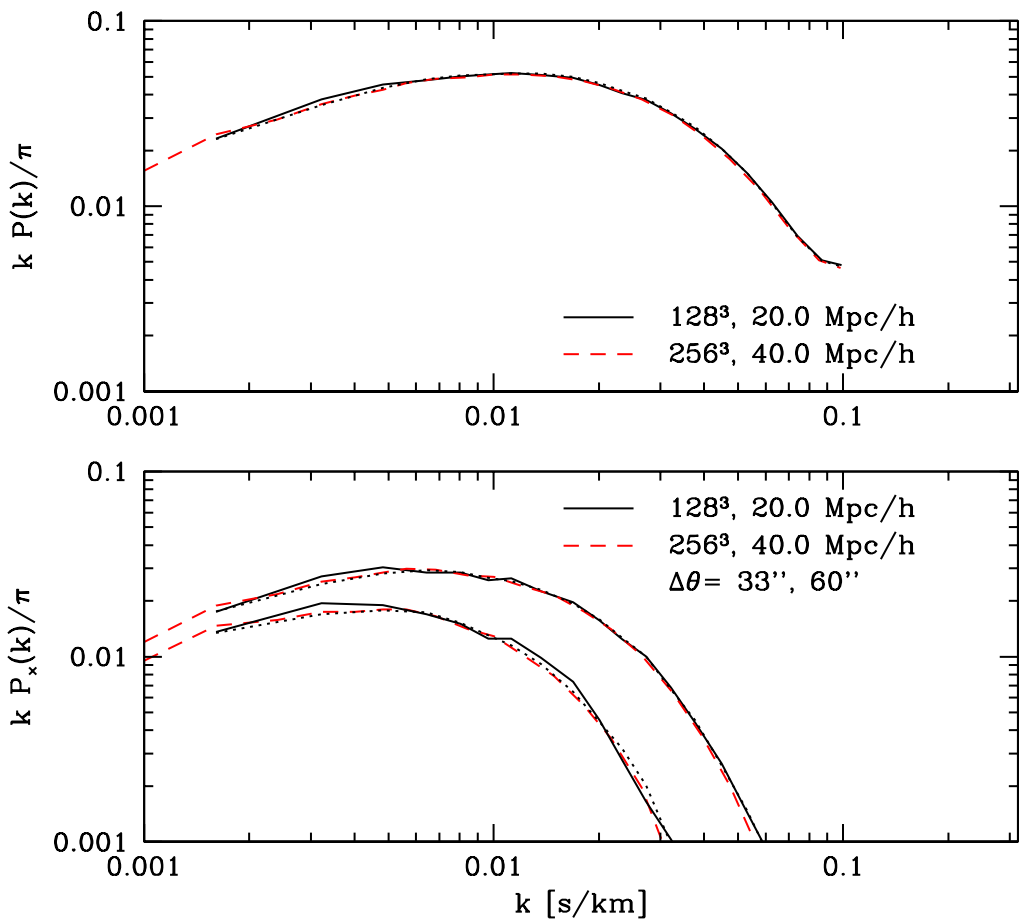,height=6.0in}}
\caption{\label{boxsize}
A boxsize test, showing the flux auto and cross spectra for a 
$128^3$ particle, $20.0$ Mpc/h boxsize simulation and those measured
from a $256^3$ particle, $40.0$ Mpc/h simulation in the same model.
The assumed IGM model is $(a, n, k_f, T_0, \alpha, \langle f \rangle) = 
(0.19, 0.7, 35.0$ h Mpc$^{-1}$, $250$ (km/s)$^2$, $0.2, 0.684)$. The
cross spectrum is computed assuming $\Omega_m=1.0$, and
shown at separations of $\Delta \theta = 33''$ and
$\Delta \theta = 60 ''$. The auto spectrum
shown here comes from the average of $6,000$ lines of sight taken along
each of the simulation box axes. The cross spectra come from the
average of $3,000$ pairs of lines of sight along each of the simulation box
axes. The black dotted
lines shows the result of averaging measurements of the auto and cross
spectrum over 5 independent simulation realizations of the same model using
the small volume simulations.
}
\end{figure}

Simulations of the Ly$\alpha$ forest are challenging in that the requirements
on box size and resolution are fairly stringent. On the smallest scales,
one must resolve the pressure smoothing scale, which is typically 
$k_f \sim 35$ h Mpc$^{-1}$ for an SCDM cosmology. 
One also needs to resolve the thermal broadening
scale which smooths the optical depth on scales of 
$k_{\rm th} \sim 0.1$ s/km, for $T_0=10^4$ K. 
On the largest scales one requires a boxsize
large enough that fluctuations on the box scale are still in the linear 
regime, and such that the box has a sufficient volume to provide a 
representative sample of the universe. In figure (\ref{resolution}) we
show a test of the resolution of our simulation box. In this figure
we show the flux auto and cross spectra from two simulations run 
in the same model, each
with a box size of $20.0$ Mpc/h, but one with $256^3$ particles and the other
with $512^3$ particles.\footnote{For the flux power spectrum, the relevant 
size for determining the resolution is that of the simulation box
in units of km/s, rather than the size in co-moving Mpc/h. Our SCDM simulation
should thus be roughly equivalent to a LCDM simulation with a box size of
$20/\sqrt{3} \sim 37$ Mpc/h.} Right at the fundamental mode, probably 
due to the small number of modes available in the simulation box, the
auto and cross spectra from the $512^3$ simulation are substantially larger
than those from the $256^3$ simulation, 
otherwise the spectra from the low and high resolution simulations are
consistent on large scales. In comparing the observed auto spectrum with
simulation models we thus neglect scales with $k < 2 \times k_{\rm fun}$,
where $k_{\rm fun}$ is the fundamental mode of the simulation box.
In comparing the cross spectrum with simulation models we include these
modes, which should be adequate since the error bars on the cross spectrum
are very large on such scales.
On small scales the $256^3$ simulation clearly
misses small scale flux power in the auto spectrum, primarily at 
$k \gtrsim 0.03$ s/km. Since we rely in this paper on
$256^3$ simulations, we then only include data with $k \lesssim 0.03$ s/km 
in comparing our models with observations, as mentioned 
in \S \ref{sim_method}. 
The cross spectrum, at separation
$\Delta \theta = 33''$, on the other hand, is actually slightly 
larger on small scales 
in the $256^3$ simulation than in the $512^3$ simulation. In this paper,
our data have poor resolution, (see Table \ref{pairtab} and the dotted lines
in the figure), so we do not use the high $k$ simulation 
points in comparing with data anyway.
In figure (\ref{boxsize}) we show a box size test. The figure shows 
a comparison between cross and auto spectra
measured in a $128^3$ particle, $20.0$ Mpc/h box size simulation and those
measured in a $256^3$ particle, $40.0$ Mpc/h box size simulation. The two
simulations thus have the same resolution, but one has twice the box size
of the other. Additionally we show the auto and cross spectra in the smaller
volume simulation using the same model, but averaged over 5 different 
realizations of the simulation box. On small scales the auto and cross spectra
are quite similar. On large scales one can see the sample variance error from
having only one realization of the smaller box. The figure also shows that
this error is quite reduced by averaging over 5 different simulation 
realizations of the smaller volume box. In conclusion, it is less
than ideal to obtain constraints from a single realization of a $256^3$,
$20.0$ Mpc/h simulation box. Especially as the 
error bars on the observed auto and cross
spectra decrease, it will be necessary to simulate a larger volume at higher
resolution.

\newpage

\newpage

\newpage

\newpage

\newpage

\newpage
\begin{deluxetable}{lcccc}
\tablecaption{\label{pairtab} 
Pairs for Cross-Correlation Analysis}
\tablehead{
\colhead{QSO Pair} &
\colhead{$z_1$, $z_2$}  &
\colhead{$\lambda_{\rm low}$(\AA) -  $\lambda_{\rm high}$(\AA)} &
\colhead{$\bar z_{Ly \alpha}$} &
\colhead{$\Delta \theta$ (arcsec)} 
}
\startdata
Q2139-4504A/B     &  3.06, 3.25	&   4467.56 -- 4811.62 & 2.82 & 33  \nl
Q2139-4433/34     &  3.22, 3.23 &   4456.54 -- 5000.52 & 2.89 & 62  \nl
KP76/77           &  2.467, 2.521 & 3809.60 -- 4108.31 & 2.26 & 147 \nl
KP76/78           &  2.467, 2.607 & 3809.60 -- 4108.31 & 2.26 & 130 \nl
KP77/78           &  2.521, 2.607 & 3809.60 -- 4108.31 & 2.26 & 180 \nl
\enddata
\tablecomments{
A summary of the quasar pairs included in the
cross-correlation analysis. The first column gives the names of the 
members of the QSO pair, the second column gives the redshifts of
each member of the pair, the third column shows the wavelength range
analyzed, the fourth column indicates the mean redshift of absorption in the
Ly$\alpha$ forest, and the fifth column is the separation of the pair
in arcseconds. The pair Q2139-4433/34 has a typical resolution
of FWHM = $3.6$ \AA, the pair Q2139-4504A/B has a typical resolution of
FWHM = $3.0$ \AA, and the KP triplet has a typical resolution of
FWHM = $2.79$ \AA.}
\end{deluxetable}

\newpage

\newpage

\newpage

\newpage

\newpage

\newpage

\newpage

\newpage

\newpage

\newpage
\end{document}